\documentclass[twocolumn,showpacs,preprintnumbers,amsmath,amssymb]{revtex4}
\usepackage{graphicx}% Include figure files
\usepackage{dcolumn}% Align table columns on decimal point
\usepackage{bm}% bold math

\begin{document}

\title{Temperature dependence of the bulk energy gap in underdoped Bi-2212:
Evidence for the mean-field superconducting transition.}

\author{V. M. Krasnov}
\email{Vladimir.Krasnov@physto.se}

\affiliation{Department of Physics, Stockholm University, AlbaNova
University Center, SE-10691 Stockholm, Sweden}

\date{\today }

\begin{abstract}

Understanding of the puzzling phenomenon of high temperature superconductivity requires reliable spectroscopic information about temperature dependence of the bulk electronic density of states.
Here I present a comprehensive analysis of $T-$evolution of bulk electronic spectra in Bi-2212 obtained by Intrinsic Tunneling Spectroscopy on small mesa structures. Unambiguous spectroscopic information is obtained by obviation of self-heating problem and by improving the spectroscopic resolution. The obtained data indicate that the superconducting transition maintains the mean-field character down to moderate underdoping, and is associated with an abrupt opening of the superconducting gap, which is well described by the conventional BCS $T-$dependence. The mean-field critical temperature reaches maximum at the optimal doping and decreases with underdoping. Such behavior is inconsistent with theories assuming intimate connection between superconducting and antiferromagnetic spin gaps, and support proposals associating high temperature superconductivity with the presence of competing ground states and a quantum critical point near optimal doping.


\end{abstract}

\maketitle

\section{Introduction}

How does high temperature superconductivity (HTSC) evolve with decreasing temperature and what happens at the superconducting transition? Where is the real critical temperature $T_c$?
Does HTSC becomes stronger or weaker upon approaching the undoped antiferromagnetic state? These highly debated questions are crucial for understanding the puzzling HTSC phenomenon. The answers to all those questions could be obtained from the analysis of temperature dependence of the superconducting energy gap $\Delta(T)$ in the quasiparticle (QP) density of states. 

So far the majority of spectroscopic studies on HTSC were made by surface sensitive techniques \cite{Ding,ARPESp,Lee2007,ARPESreview,Renner,Yazdani,Misra}. However, obtaining reliable spectroscopic information from surface spectroscopy on HTSC is immensely difficult: atomic-scale $c-$axis coherence length, rapid chemical deterioration, presence of the surface states \cite{ARPESreview} and inherently different doping state of the surface, may preclude determination of bulk electronic properties by surface sensitive techniques. Furthermore, for Bi$_2$Sr$_2$CaCu$_2$O$_{8+\delta}$ (Bi-2212) the surface spectroscopy probes the BiO rather than the superconducting CuO$_2$ plane \cite{Misra}. All this urges the necessity of bulk spectroscopy of HTSC.

Intrinsic Tunneling Spectroscopy (ITS) provides a unique opportunity to probe bulk electronic properties of HTSC. This relatively new technique utilizes weak interlayer ($c-$axis) coupling in quasi two dimensional HTSC compounds, in which mobile charge carriers are confined in CuO$_2$ planes separated by some blocking layer (e.g. SrO-2BiO-SrO in case of Bi-2212). This leads to formation of natural, atomic scale intrinsic tunnel junctions, and to appearance of the intrinsic Josephson effect at $T<T_c$ \cite{Kleiner,Fiske,LatyshPhC,Kim,Wang,Fluct,MQT,Latysh,Ooi}.
ITS is well suited for clarification of questions highlighted above: it is a direct spectroscopic technique, is not sensitive to phase coherence, has very high resolution ($neV$ achievable), is mechanically stable and thus perfectly suited for $T-$dependent studies of HTSC (unlike surface probe techniques), and, most importantly, probes bulk electronic properties of HTSC.

This work represents a comprehensive analysis of temperature dependence of the bulk energy gap obtained by ITS on small Bi-2212 mesa structures. Unambiguous $\Delta(T)$ is obtained by careful
obviation and cancelation of self-heating. Improved resolution by means of $T-$differential ITS allows tracing the gap in the phase-incoherent state at $T\gtrsim T_c$. It is observed that at all studied doping levels, the superconducting gap opens abruptly in a mean-field manner and is well described by the conventional BCS temperature dependence. The mean-field critical temperature $T_c^{mf}$ decreases with underdoping, thus confronting speculations about persistence of superconductivity up to very high temperatures above $T_c$ in underdoped HTSC. In most underdoped crystals a remaining, weakly $T-$dependent pseudogap (PG, $\Delta_{PG}$) is observed at $T_c^{mf} <T <T^*\sim 120-150 K$. The pseudogap seems to form a combined gap with the BCS-like superconducting gap below $T_c^{mf}$, and thus compete with superconductivity. No signature of the PG is observed at the optimal doping. The obtained results are inconsistent with theories assuming intimate connection between superconducting and antiferromagnetic spin gaps, and support proposals associating HTSC with the presence of competing ground states and a quantum critical point near optimal doping.

In conventional low temperature
superconductors (LTSC) superconducting transition occurs as a result of the second-order phase transition. It is associated with abrupt appearance of the order parameter, represented by the superconducting energy gap $\Delta(T\lesssim T_c^{mf}) \propto \sqrt{1-T/T_c^{mf}}$,
and by linear growth of the upper critical field: $H_{c2} \propto \Delta^2 \propto 1-T/T_c^{mf}$. The correlation $H_{c2} \propto \Delta^2$ is fundamental, because $H_{c2}$ is inversely proportional to the square of the coherence length $\sim$ Cooper pair size, which is inversely proportional to the pair coupling energy $\Delta$. All this is perfectly described by the mean-field BCS-Eliashberg theory of superconductivity \cite{BCSE}.

But how does HTSC emerge with decreasing temperature, and what happens at $T_c$ remain unclear. For overdoped cuprates, thermodynamic characteristics reveal unambiguous evidence of the second-order phase transition at $T_c$ \cite{TallonPhC}. Similarly, analysis of Nernst effect \cite{Huebener}, equilibrium \cite{Landau} and fluctuation \cite{Fluct1,Fluct2} magnetization and resistivity \cite{Bouquet} reveal vanishing of $H_{c2}(T\rightarrow T_c)$ in a wide doping range. However, in underdoped cuprates, characterized by the persistence of the normal state pseudogap at $T>T_c$, the superconducting transition at $T_c$ becomes obscured and both thermodynamic \cite{TallonPhC} and transport \cite{Ong} properties become abnormal. According to some reports neither $\Delta$ \cite{Renner}, nor $H_{c2}$ \cite{Ong} vanish at $T_c$, although different interpretations of similar data are possible \cite{Fluct1,Yazdani}, and vanishing gap at $T \rightarrow T_c$ was also reported  \cite{Deutscher,KrTemp,Doping,Raman,Lee2007}.

A related controversy exists about doping dependence of the coupling strength in HTSC. Although $T_c$ and $H_{c2}$ decrease with underdoping \cite{Hc2,Ong}, the gap measured by surface sensitive techniques was reported to grow \cite{Renner,Ding}. This has been taken as evidence for a continuously increasing superconducting coupling strength $\propto \Delta/T_c$ upon approaching the antiferromagnetic state, assuming an intimate connection between the two states \cite{AF}. If true, this would indicate that HTSC has a magnetic origin. However, other experiments reveal the existence of two distinct energy scales, of which one indeed increases with approaching the antiferromagnetic state, while the other follows $T_c$ at all doping levels \cite{Deutscher,KrTemp,Doping,Raman,Lee2007}.

Does HTSC becomes stronger or weaker with approaching the antiferromagnetic state? Again, the answer could be obtained by understanding, what happens at $T_c$.
If the coupling strength is increasing with underdoping, then so does $T_c^{mf}$. Some researchers assume that $T_c^{mf}$ may exceed the room temperature already at moderate underdoping \cite{Renner,Ong,Tesanovic}. To cope with the apparent decrease with underdoping of transport $T_c^{phase}$, at which phase coherence is achieved in transport measurements, one has to assume the existence of an extended region $T_c^{phase}<T<T_c^{mf}$ in which the amplitude of the superconducting order parameter is large but the phase coherence is destroyed by thermal fluctuations \cite{Precursor}.

The extent of the phase incoherent state $T_c^{mf}-T_c^{phase} \sim G_i T_c^{mf}$ is described by the Ginzburg-Levanyuk parameter $G_i$ \cite{Larkin}. For clean LTSC the fluctuation region is very small because of very small $T_c^{mf}/T_F \sim 10^{-4}$, where $T_F$ is the Fermi temperature \cite{Larkin}. Even for HTSC, $T_c^{mf}/T_F < 0.1$. Therefore, expansion of the phase-incoherent state well below $T_c^{mf}$ requires $G_i \sim 1$, which can be achieved only by decreasing the dimensionality of the system \cite{Larkin}. Thus one has to assume that superconductivity is either one-dimensional (e.g. due to the presence of stripes \cite{AF}) or zero-dimensional (as in granular superconducting films \cite{Granular}). In this case, there would be no second order phase transition, nor significant amplitude fluctuations of the order parameter upon establishing of the phase coherence at $T_c^{phase}$, the superconducting gap would persist at $T_c^{phase}<T<T_c^{mf}$, and could be directly measured by tunneling spectroscopy. Therefore, the knowledge of $\Delta(T)$ close and above $T_c^{phase}$ is crucial for understanding HTSC.

An important clue to understanding temperature evolution of electronic states in HTSC was provided by recent surface photoemission experiments \cite{ARPESp}, which, unlike earlier works, showed that $\Delta$ does not have a simple $d-$wave momentum dependence, but is described by two distinct energy scales \cite{Lee2007}. The anti-nodal gap has weak temperature dependence at $T \lesssim T_c$ and turns into the pseudogap at $T>T_c$. Furthermore, it increases with underdoping and tends to merge with the insulating gap in the undoped antiferromagnetic state. On the other hand the gap in the nodal ``Fermi arc" region must be associated with superconductivity because it follows $T_c$ and vanishes close to $T_c$ at all doping levels. Although those observations are consistent with several previous reports \cite{KrTemp,Doping,Raman}, the reliability of surface spectroscopy of HTSC is now under question, because it fails to reveal the electron-like Fermi surface, uncovered by recent quantum oscillation experiments \cite{VanAlphen}.

The paper is organized as follows.
In Sec. II I describe samples and emphasize the dramatic difference between surface and bulk properties of Bi-2212 crystals. Sec. III contains basic characterization of temperature and size-dependencies of ITS on small mesas. In sec. IV main experimental results on temperature dependence of the bulk gap in Bi-2212 are presented. It is shown that self-heating is effectively obviated by decreasing mesa size and can be simply canceled out from experimental data. The observed results are discussed and summarized in sec.V. In Appendix-A, self-heating and non-equilibrium effects in intrinsic tunnel junctions are analyzed. %It is discussed how self-heating distorts intrinsic tunneling characteristics. It is shown that self-heating phenomenon is trivial (perhaps the simplest as far as HTSC spectroscopy is concerned), is obvious when present and can be easily obviated and compensated.
In Appendix-B artifacts of in-plane resistance and limitations on the junction size are discussed. %of it is demonstrated that acute back-bending can occur in large structures even without heating. The experimentally observed difference between 4-and 3-probe measurements is also explained.

\begin{table*}
%\noindent
%\begin{minipage}{.95 \textwidth}
\caption{Parameters of studied samples: $N-$ number of IJJs; $T_c^{onset}$ - onset of the resistive transition; $T_c^{phase}$ - appearance of the measurable critical current and establishing of the $c-$axis phase coherence; $T_c^{mf}$- mean-field critical temperature, obtained from extrapolation of $\Delta (T)$;
$\Delta_{SG}(0)$ the superconducting gap at $T\rightarrow 0$; $\Delta_{PG}$ the pseudogap at $T_c^{mf}$; $U_{TA}$- the $c-$axis thermal activation barrier in the normal state; $J_c(0)$- the critical current density at $T\rightarrow 0$.
All samples were made from the same batch of  Bi$_2$Sr$_2$Ca$_{1-x}$Y$_x$Cu$_2$O$_{8+\delta}$ single crystals. }
%\texttt{table}
%\textbackslash\texttt{multicolumn}
\begin{ruledtabular}
\begin{tabular}{cccccccccc}
%&\multicolumn{2}{c}{}&\multicolumn{2}{c}{}&\\
Sample&$N$&$T_c^{onset}$&$T_c^{phase}$&$T_c^{mf}$&$\Delta_{SG}$(0)& $\Delta_{PG}$ &$U_{TA}$&$J_c(0)$&References \\

 & & (K) & (K) & (K) & (meV) & (meV) & (meV)&(A/cm$^2$) & \space \\
\hline

SMa &9& 95 & 91.5 & 96 & 33.9 & 0& 25 & 1000 & Ref.\cite{Heat}\\
\hline

S42& 9 & 93.3 & 92.1 &93& 33.4 & 0& 22 & 1100 & Ref.\cite{KrTemp}\\
\hline

S92 & 34 & 86 & 80 & 90 & 34.6 & 0& 34 & 300 & Refs. \cite{Heat,Fluct,Sven} \\
    &    &    &  $or$  &87 & 32.7 &12 &    & & \\
\hline

S43 & 8 & 86.5 & 81.5 &90& 42.5 &0 & 32.5 & 510 & Ref.\cite{Doping} \\
&&&$or$&85& 38&19&&&\\
\hline

S82 & 7 & 78 & 73 &89& 46 &0& 28 & 270 & Ref.\cite{Doping} \\
&&&$or$&87&44.8&11&&&\\
%\hline
%S512 & 8 & 91 & 86 & 87 & 27.5 &0& 19.5 & 650 & Annealed at 600C, Ref.\cite{Doping,Cascade} \\

\end{tabular}
%\end{minipage}
\end{ruledtabular}
\end{table*}

\section{Experimental}

Interlayer tunneling occurs in various layered HTSC compounds, such as Bi-2212 \cite{KrTemp,Doping,Suzuki,Latysh,Lee,Cascade}, Bi-2201 \cite{YurgensBi2201}, Tl-2212\cite{Schlen,Warburton} YBaCuO \cite{YBaCu}, and some others \cite{RuCuO}, as well as in intercalated compounds \cite{KrMag}. ITS was also expanded to non-HTSC layered compounds \cite{LatyshevNbSe,Manganite}.

Observation of the intrinsic Josephson effect \cite{Kleiner} at $T<T_c$ provides the most clear evidence for interlayer tunneling in strongly anisotropic layered HTSC. At present all major fingerprints of
the intrinsic Josephson effect were observed, including Fiske
\cite{Fiske,LatyshPhC,Kim} and Shapiro \cite{LatyshPhC,Wang} steps
in Current-Voltage characteristics (IVC); the Josephson plasma
resonance \cite{Plasma}; thermal activation \cite{Fluct} and
macroscopic quantum tunneling \cite{MQT} from the Josephson
washboard potential; and the flux quantization
\cite{Fiske,LatyshPhC,Kim,Latysh,Ooi}. The latter experiments
explicitly confirmed the correspondence between the stacking
periodicity of intrinsic Josephson junctions (IJJs) and the
crystallographic unit cell of Bi-2212.

Several techniques for preparation of IJJs have been developed, such as patterning mesa structures on top of single crystals \cite{YurgensPRL}, 3D-sculpturing by Focused Ion Beam (FIB) \cite{Kim,Latysh} and double-side fabrication \cite{Wang}.

Here all measurements were performed on small mesa structures because they are best suited for ITS: they have the best thermal anchoring and are less prone to self-heating and other artifacts, as described in the Appendix. All measurements were made in the three-probe configuration, which is more robust toward artifacts, as discussed in Appendix-B.

\subsection{Samples}

To avoid variations caused by different crystal stoichiometry, single crystals from the same batch of Y-substituted Bi$_2$Sr$_2$Ca$_{1-x}$Y$_x$Cu$_2$O$_{8+\delta}$ (Bi(Y)-2212) were used in this work, with the onset of superconductivity at optimal doping at $T_c^{onset}\simeq 96$K. Obtained results are not specific for this batch, see e.g. Ref.\cite{Suzuki,Lee,Sven}.

Several mesas of sizes from $\sim 7\times 7$ to $\sim 3\times 3 \mu m^2$ were patterned simultaneously on each crystal by photolithography and Ar-ion milling. % or wet chemical etching. 
All mesas usually contained the same amount of IJJs $N$. The virgin crystals were slightly overdoped with $T_c \sim 91K$. %Wet etching does not involve major thermal treatment of crystals and preserves the original doping level in mesas, but at expense of poor spatial control of etching. To the contrary, 
Ar-ion milling, provides very uniform and controlled etching, but is accompanied by substantial heating at high-vacuum. This results in partial out-diffusion of Oxygen, so that mesas become underdoped. %Some crystals (sample S512) were annealed in vacuum at 600 C for several hours prior to mesa fabrication. Although the final $T_c$ of such mesas was approximately the same, the $c-$axis resistivity and gaps were somewhat lower than in unannealed mesas. This difference, referred to as the "small" and "large" gap cases in Ref.\cite{Doping}, probably reflects different doping inhomogeneity, with annealed crystals presumably being more homogeneous. 
To reduce self-heating, some mesas were trimmed to sub-micron size by FIB. Details of mesa fabrication are described in Ref.\cite{Fluct}.

Table-I summarizes crystals used in this study. To save space and avoid repetitions, I show only a limited amount of raw data and instead address the reader to previous works listed in Table-I.
%:
%the sample SMa was used for analysis of size dependence of ITS characteristics, see Fig. 2 in Ref. \cite{Heat}; S42 was studied in Ref.\cite{KrTemp}; data for S512 were shown in Fig. 1c) of Ref.\cite{Doping} and Fig.2 b,d) from Ref.\cite{Cascade}; $dI/dV(V,T)$ curves for a mesa from S43 were shown in Fig. 3 from Ref.\cite{Doping}; the sample S92 was used for in-situ measurement of self-heating, see Fig. 3 in Ref.\cite{Heat}, periodicity and scaling of QP branches in IVC's for mesas of different areas from S92 was demonstrated in Fig.2 of Ref.\cite{Fluct} and Fig.2 of Ref.\cite{HeatCom}, $dI/dV(V,T)$ curves for the smallest mesa 4aF2 from S92, which is analyzed below,  were shown in Fig.2 of Ref.\cite{Sven}; IVC for a mesa from S82 was shown in Fig. 1d) from Ref.\cite{Doping}. 
All of the studied mesas exhibited good periodicity of QP branches in the IVC, indicating good uniformity of IJJs in the mesas (see the insets in Fig. 2a) and references in Table-I).

The solid line in Fig. 1 a) shows a typical resistive transition for a near optimally doped mesa on SMa sample. Here $R_0$ is the zero bias resistance measured with small $\sim 1 \mu A$ ac current. One can see two transitions (three branches) in $R_0(T)$ : the major part of the mesa goes into the superconducting state at $\sim 92K$, while the final transition takes place at $\sim 40K$. To clarify their origins, the IVC of this mesa at $4.7K$ is shown in the inset of Fig. 1 a). It is seen that the IVC is strongly hysteretic and consist of periodic QP branches due to sequential switching of IJJs between superconducting and resistive states. The hysteresis is typical for tunnel junctions and is caused by low damping and large specific capacitance of the junctions \cite{Collapse,CollapsePRB}, which allows junctions to remain in the resistive state even below the critical current.
Distinct branches in $R_0(T)$ in the superconducting state are originating from different sections of the hysteretic IVC, as indicated by arrows in Fig. 1 a).

Experiments on small mesas allow measurement of the QP
resistance at different bias, $R^{QP}(I)$, in the superconducting state \cite{YurgensPRL,Latysh,Sven}, which is otherwise shunted by the supercurrent \cite{Morozov}. Circles in Fig. 1 a) show the zero-bias QP resistance $R_0^{QP}$ obtained by extrapolation to $I\rightarrow 0$ of the last QP branch with all IJJ's in the resistive state. The QP resistance at different bias can be also measured explicitly by first pulsing a current above the critical current $I_c$ and then ramping it down to a desired value \cite{Sven}. Bias yields an additional parameter for intrinsic tunneling studies, which may render crucial for correct interpretation of the data \cite{Sven}.

\subsection{Surface versus bulk properties}

As seen from the inset in Fig.1a), the critical current of the first junction $\sim 20\mu A$ is much smaller than that for the rest of the junctions $I_c\sim 270 \mu A$. This is the top IJJ in the mesa, between the two outmost CuO planes. It is seen that the surface CuO plane is superconducting, but has a lower critical temperature $T_c'\simeq 40K$.

%The question whether or not the $T_c'$ and, consequently, the electronic properties of the surface are fundamentally different from that of the bulk is important because so far the majority of spectroscopic information on HTSC was obtained by surface photoemission \cite{ARPESp,Lee2007,ARPESreview} and surface tunneling \cite{Renner,Yazdani} techniques. Data in Fig. 1a) clearly demonstrates that the surface of Bi-2212 can be totally different from the bulk.
Noticeably, the critical current and the QP resistance of the second junction, formed by the second and third CuO planes below the surface, is practically the same as for the rest of the junctions. This unambiguously shows that suppression of superconductivity is solely the surface phenomenon and occur only in the top CuO plane, and that ``bulk" behavior starts already from the second CuO layer below the surface.

\begin{figure}
\includegraphics[width=3.0in]{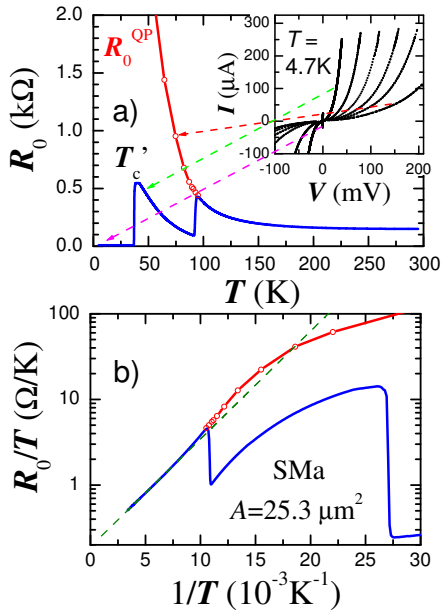}
\caption{\label{Fig1} (Color online). a) Zero bias resistance $R_0$ vs $T$ for a mesa on a nearly optimally doped crystal SMa. Inset shows $I-V$ characteristics at $T=4.7K$ which clarifies the origin of certain parts of the resistive transition. b) The same data shown as a thermal activation plot $R_0/T$ (in a logarithmic scale) vs. $1/T$. It is seen that in the whole normal state region $T>T_c$, $R_0$ is described by the Arrhenius law (dashed line) with a constant TA barrier $U_{TA}\simeq 25 meV$.}
\end{figure}

The reduction of surface $T_c'$ in our mesas is predominantly caused by chemical deterioration in atmosphere during a short period between cleavage of the Bi-2212 crystal and deposition of the top Au protection layer. Such deterioration was studied in detail in Ref.\cite{Wei}, where it was shown that $T_c'$ could be increased to $\lesssim 80K$, if cleavage and deposition are made quickly without breaking vacuum (replicating conditions for surface spectroscopy of HTSC \cite{ARPESp,Lee2007,ARPESreview,Renner,Yazdani} ). This is still substantially lower than the bulk $T_c$.

\begin{figure}
\includegraphics[width=2.8in]{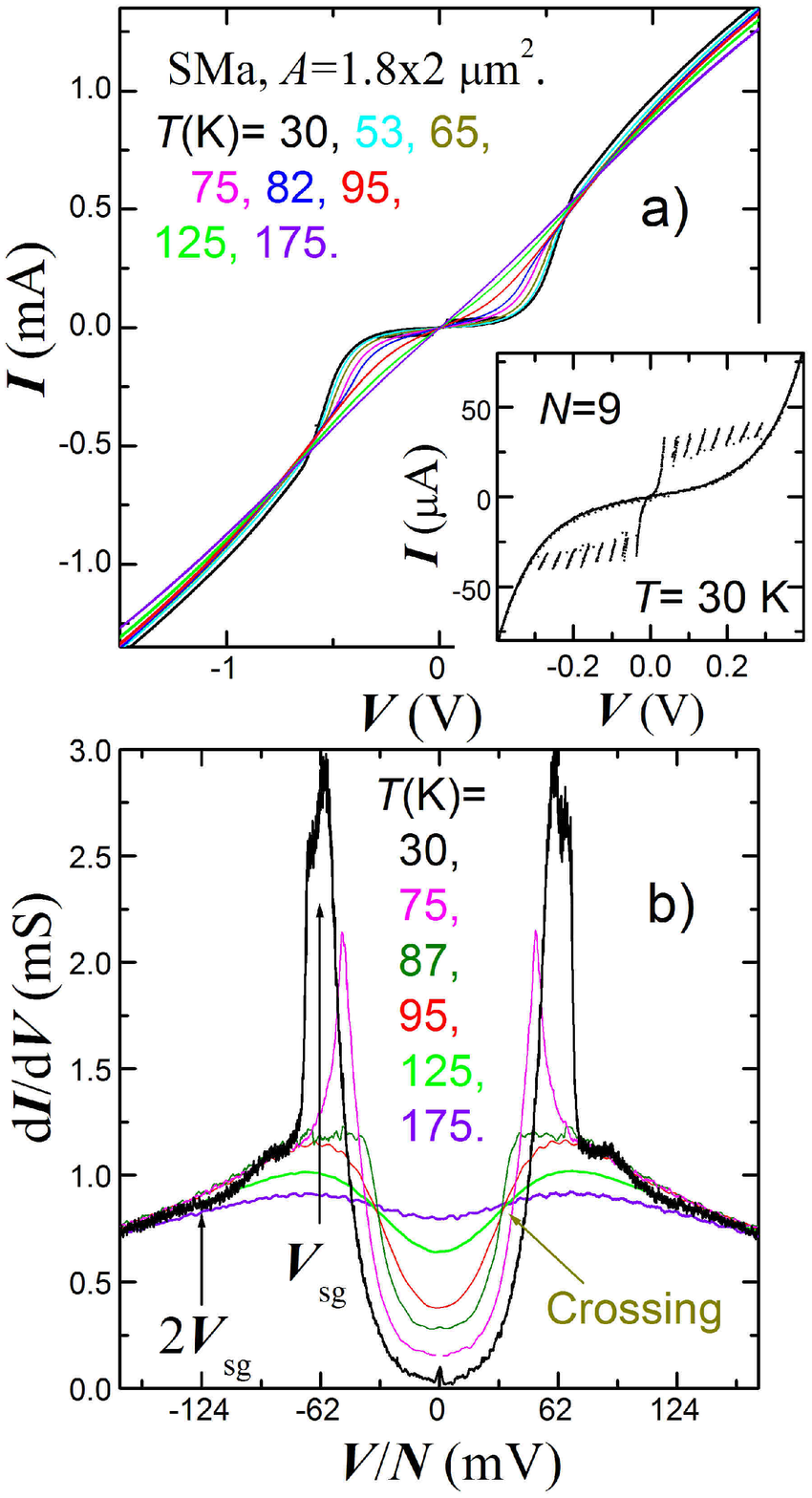}
\caption{\label{Fig2} (Color online). a) $I-V$ curves at different $T$ for a small mesa on a near optimally doped SMa crystal. The kink and transition to ohmic resistance at the sum-gap voltage is clearly seen. Inset shows multiple quasiparticle branches due to one-by-one switching of intrinsic Josephson junctions into the resistive state. b) $dI/dV$ vs. voltage per junction for the same mesa. Arrows indicate the following characteristic features: the sum-gap peak, a minor double-gap dip at $T<T_c$, and a crossing point at $T>T_c$.}
\end{figure}

The remaining suppression of $T_c'$ of the surface layer is often attributed to the proximity effect with the electrode. However, I would like to note that although the top CuO layer has a lower $T_c'$ (in some samples less than 20K) the second layer is perfectly ``bulk", i.e. not deteriorated with respect to deeper laying layers. Thus, there is no detectable proximity effect between the first and the second CuO layers. This is natural because the transparency of the interlayer barrier is low (prerequisite of the tunnel junction) and the $c-$axis coherence length is sub-atomic ($<0.3 {\AA}$). Exactly for the same reasons there should be no considerable proximity effect between the electrode and the top CuO layer, because those are also separated by the blocking BiO layer.

More likely, the suppressed $T_c'$ reflects the fundamental difference between the surface and the bulk e.g., because the surface is lower doped than the bulk. It should be also noted that even in UHV conditions the chemical deterioration of un-protected Bi-2212 surface is non-negligible. Note that the time of deposition of a monolayer is only $\sim 20 s$ at the residual pressure $p=10^{-7}Torr$ and even in state of the art surface spectroscopy systems is at best a matter of few hours.

In any case the observed abrupt transition from the surface to the bulk properties within just one atomic layer from the surface clearly indicates that it is not at all granted that the surface spectroscopy can uncover substantial information about bulk electronic properties of HTSC. All this urges the necessity of bulk spectroscopy of HTSC, as emphasized in the Introduction.

\section{Intrinsic tunneling characteristics of small mesas}

Figs. 2 a) and b) show the $c-$axis $I-V$ and $dI/dV$ characteristics of a small mesa on the near optimally doped SMa sample at different $T$. A pronounced kink in $I-V$ (peak in $dI/dV$), followed by the ohmic and almost $T-$independent tunnel resistance is seen \cite{KrTemp}. Such the IVC is typical for superconducting tunnel junctions, as demonstrated in Fig. 3 a), and is associated with the sum-gap singularity at $V_{sg}= 2\Delta/e$, providing the basis for ITS and opening a possibility to study bulk electronic spectra of HTSC.

\begin{figure}
\includegraphics[width=3.5in]{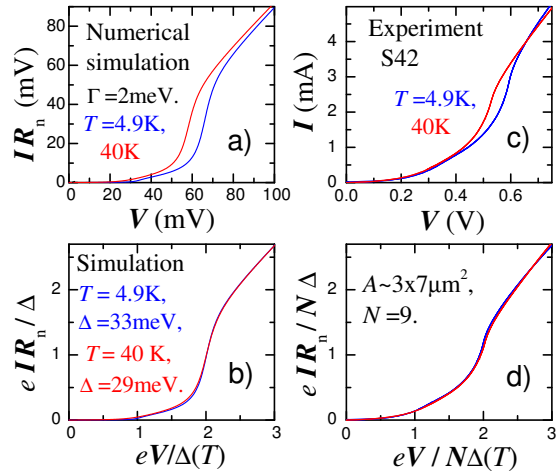}
\caption{\label{Fig3} (Color online). Comparison of a) simulated and c) experimental IVCs for S42  at 4.9 K and 40 K; b) and d) demonstrate collapse of the $I-V$ curves, scaled by $\Delta(T)$ and indicate that both $V-$ and $I-$ scales are determined by $\Delta(T)$.}
\end{figure}

Fig. 3 a) shows numerically simulated IVCs at $T=$4.9K and 40 K for a superconducting tunnel junction with a gapless density of states at the Fermi level. The gaplessness was achieved by introducing an appropriate depairing factor $\Gamma = 2 meV$ into the BCS density of states. Parameters were chosen to fit experimental data for the S42 sample, shown in Fig. 3 c). Clearly, $\Delta$ decreased from 33 meV at 4.9 K to 29 meV at 40 K. Fig.3 b) shows the same calculated IVC's in which both current and voltage scales are normalized by the gap. It is seen that the curves merge, because at low $T \lesssim T_c/2$ both $V-$ and $I-$ scales $\propto \Delta$. A similar scaling is also observed for experimental data, shown in Fig. 3 d), confirming that the kink is related to the sum-gap singularity. Additional arguments can be found in the Appendix A-1.

\begin{figure}
\includegraphics[width=3.0in]{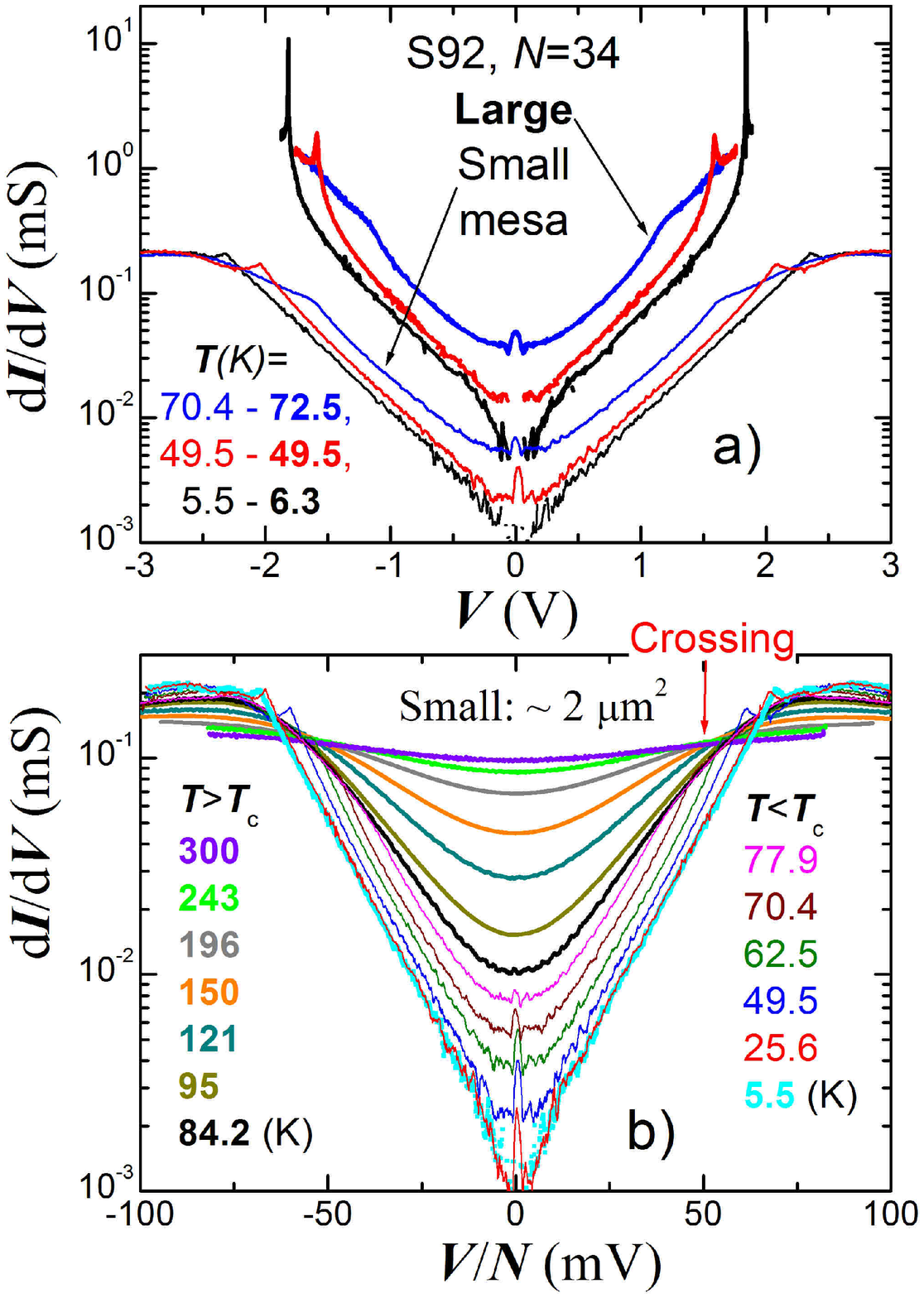}
\caption{\label{Fig4} (Color online). a) $dI/dV(V)$ (in a semi-log scale) for two mesas with different area on a moderately underdoped S92 crystal. It is seen how heating and in-plane resistivity bend upwards curves at high bias and reduce the sum-gap peak voltage for the large mesa. b) $dI/dV(V)$ curves at different $T$ for the small mesa. In is seen that the curves maintain the linear V-shape (in the semi-log scale) in the whole sub-gap region when self-heating becomes negligible. A sudden crossover at $T=T_c$ from tunneling-like with $T-$independent slope, to thermal activation behavior with $T-$dependent slope is clearly seen. Data from Ref.\cite{Sven}.}
\end{figure}

The sum-gap singularity is not the only gap-related feature in ITS.
Further check for self-consistency of our interpretation can be obtained from
analysis of additional, more subtle gap-related features in $dI/dV$. The arrow in Fig. 2 b) indicates a small dip at $2V_{sg}$. It was studied in Ref. \cite{Cascade} and was attributed to enhancement of non-equilibrium effects at $eV>4\Delta$, when relaxation radiation of tunneled quasiparticles becomes sufficient for breaking Cooper pairs.

\subsection{Self-heating free characteristics}

It is fair to say that ITS has became a spectroscopic tool as a result of reduction of self-heating. No $T-$independent ohmic tunnel resistance, as in Fig. 2 a), could be seen in earlier works \cite{Kleiner,YurgensPRL,Gough}, dealing with large structures. Larger self-heating in such structures leads to development of an acute thermal instability \cite{HotSpot} at voltages much smaller than $V_{sg} N$. Another problem in large structures is associated with the limited amount of supercurrent that the CuO plane can carry (see Appendix-B). It is interesting to note that exactly the same problems were encountered at the early stage of experimental studies on LTSC tunnel junctions \cite{Giaever4p}.

\begin{figure}
\includegraphics[width=2.8in]{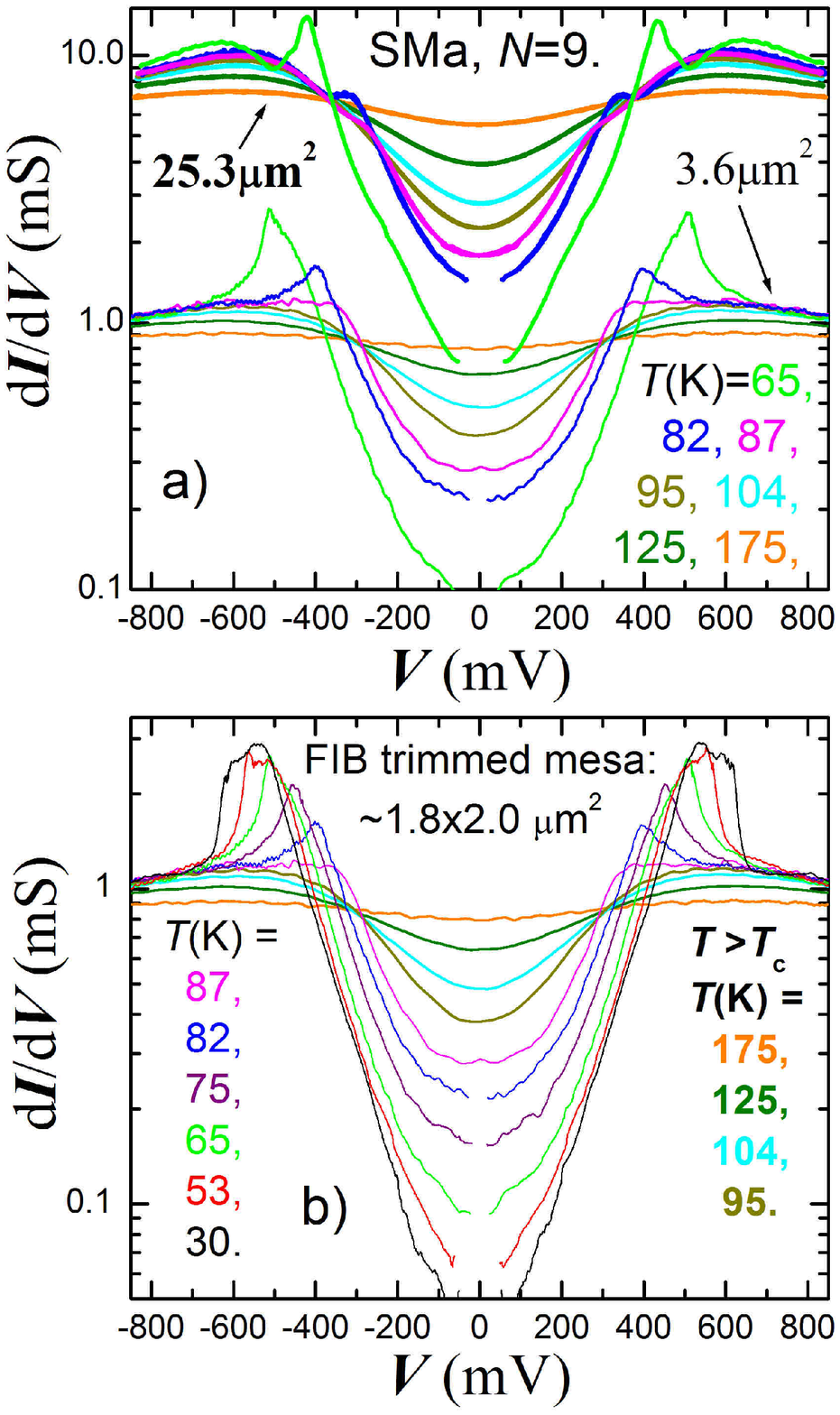}
\caption{\label{Fig5} (Color online). a) $dI/dV(V)$ (in a semi-log scale) at different $T$ for the same mesas on near optimally doped crystal SMa before and after FIB trimming. The peak voltage is reduced in the larger mesa due to self-heating. b) $dI/dV$ curves at different $T$ for the small mesa. The characteristic V-shape in the semi-log scale is observed at all $T$. As in Fig. 4, the crossover from $T-$independent to $T-$dependent slope occurs at $T_c$. }
\end{figure}

In order to obtain unambiguous spectroscopic information, the problem of self-heating has to be carefully addressed. The temperature rise due to self-heating is given by a simple expression \cite{CollapsePRB}
\begin{equation}\label{EqSHeat}
\delta T = P R_{Th}(T),
\end{equation}
where $P=IV$ is the dissipated power and $R_{Th}$ is the effective thermal resistance of the mesa, which is $T-$dependent and, therefore, bias dependent \cite{Heat}. Detailed analysis of self-heating in ITS, including numerical simulations of distortion of IVCs by self-heating can be found in Appendix-A.

In recent years different ways of obviating self-heating in ITS were employed, such as pulse measurements \cite{Suzuki}, miniaturization \cite{HeatJAP,Heat} and heat compensation \cite{Lee}.
In Ref.\cite{Heat} it was emphasized that miniaturization %(both the area and the number of junctions in the mesa) 
decreases self-heating at a given voltage per junction and provides an unambiguous way for discrimination of artifacts of self-heating (size-dependent) from electronic spectra (material property, size-independent).

In Figs. 4a) and 5a) $dI/dV(V)$ characteristics of large and small mesas on samples S92 (moderately underdoped) and SMa (near optimally doped), respectively, are shown. It is clearly seen how self-heating distorts the last QP branch in large mesas. At low bias, when self-heating is negligible, the characteristics of both mesas are undistorted and the $\log dI/dV$ curves remain linear and parallel, i.e., simply scale with the mesa area. However, at larger bias $\log dI/dV$ in the larger mesas starts to growth faster (super-linear) because the sub-gap conductance increases with $T$ (see Fig. 2b). The larger mesas also reach the peak earlier, i.e., at lower voltage, as a result of suppression of the gap by $T$.

Figs. 4 and 5 b) show $T-$evolution of $\log dI/dV(V)$ characteristics for the smallest mesas. It is seen that the $\log dI/dV$ characteristics for those mesas remain linear in the whole sub-gap region $V/N<V_{sg}$, indicating that they are not distorted by self-heating up to $V_{sg}$. For both mesas this was explicitly proven in Ref.\cite{Heat}: for SMa by size-independence of the peak voltage and for S92 by in-situ measurement of self-heating.

Note that heating-free ITS characteristics for both crystals have a remarkably trivial V-shape in the semi-log scale, and that the slope of the curves for both samples experience an abrupt crossover from thermal-activation (TA) like $1/T-$ dependence at $T>T_c$ to tunneling-like $T-$independent slope at $T<T_c$. As discussed in Ref.\cite{Sven}, this indicates opening of an additional quantum transport channel for Cooper pairs at $T<T_c$.

\subsection{Thermal-activation behavior in the normal state}

In Ref.\cite{Sven} it was shown that at $T>T_c$ ITS characteristics exhibit TA behavior. Up to moderately high bias they are described by a simple expression:

\begin{equation}
\frac{dI}{dV}(T,V) \propto \frac{1}{T}
\exp\left[-\frac{U_{TA}}{k_B T}\right]\cosh\left[\frac{eV}{2k_B T}\right], \label{TeffN}
\end{equation}

with a constant TA barrier $U_{TA}$. Indeed, the $\cosh$ term reproduces the rounded V-shape of $\ln [dI/dV](V)$ curves with the slope that increases as $\simeq 1/T$. The TA behavior at zero bias at $T>T_c$ is demonstrated in Fig. 1 b). %Detailed analysis of the TA behavior was presented in Ref.\cite{Sven}. 
In order to expand the TA-analysis to higher bias, we have to carefully integrate TA-current through the junction:

\begin{eqnarray}\label{Eq.Tunn}
I(V) = \int_{-\infty}^{+\infty} t_{TA}(E,V) \rho(E)\rho(E+eV) \\
\nonumber
\left[f(E)(1-f(E+eV))-f(E+eV)(1-f(E)) \right] dE.
\end{eqnarray}

Here $t_{TA}=\exp(-U(E,V)/k_B T)$ is the transition probability of the TA process, and $U$ is the effective TA barrier $U(E,V)= \min\left[(U_{TA}-E-eV/2); 0 \right]$, $\rho(E)$ are density of states in electrodes and $f(E)-$Fermi distributions.

\begin{figure}
\includegraphics[width=3.0in]{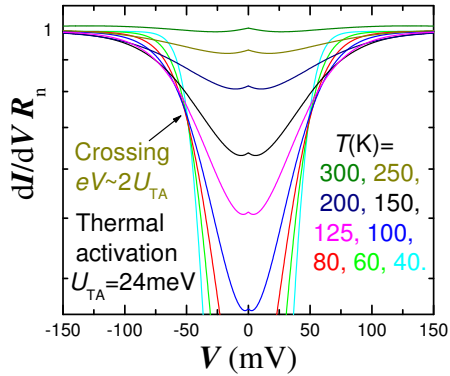}
\caption{\label{Fig6} (Color online). Simulated TA characteristics from Eq.(\ref{Eq.Tunn}) with $U_{TA}=24 meV$. They reproduce all major features of experimental curves at $T>T_c$ in Figs. 4 and 5 b) including the V-shape with the slope proportional to the reciprocal temperature and the crossing point at $eV\sim 2U_{TA}$. }
\end{figure}

Fig. 6 shows simulated $dI/dV(V)$ characteristics (in the semi-log scale) obtained from Eq.(\ref{Eq.Tunn}) with $U_{TA}=24$ meV and for {\it completely normal} electrodes $\rho(E)=$const. It reproduces all characteristic features of experimental data at $T>T_c$, including the crossing point
at $eV\simeq 2U_{TA}$ and the inverted parabolic shape at high temperatures $k_BT > U_{TA}$ \cite{Doping}.

Apparently, the $c-$axis TA barrier should be identified with the phenomenon referred to as the large $c-$axis pseudogap in the previous literature. However, the amazing success of the trivial TA model, with only one constant parameter $U_{TA}$ in the whole normal region $T>T_c$ and without any momentum dependent gap in the density of states, suggests that the $c-$axis pseudogap is most probably not the gap in electronic spectrum of CuO layers, but the property of the blocking BiO layer. Possible ``non-gap" origins of the $c-$axis TA barrier were discussed in Ref.\cite{Sven}. Those include resonant tunneling through the impurity state in the blocking layer, inelastic tunneling with excitation of a molecular mode in the barrier, and Coulomb blocking of tunneling in the poorly conducting two-dimensional electron system. As already noted in Ref.\cite{KrTemp}, the later bares a striking similarity with experimental V-shape characteristics. The Coulomb blocking depends entirely on the conductivity of the two-dimensional electron system, which would naturally explain the increase of TA barrier with underdoping.

\subsection{Improving resolution by $T-$differential spectroscopy}

\begin{figure}
\includegraphics[width=3.0in]{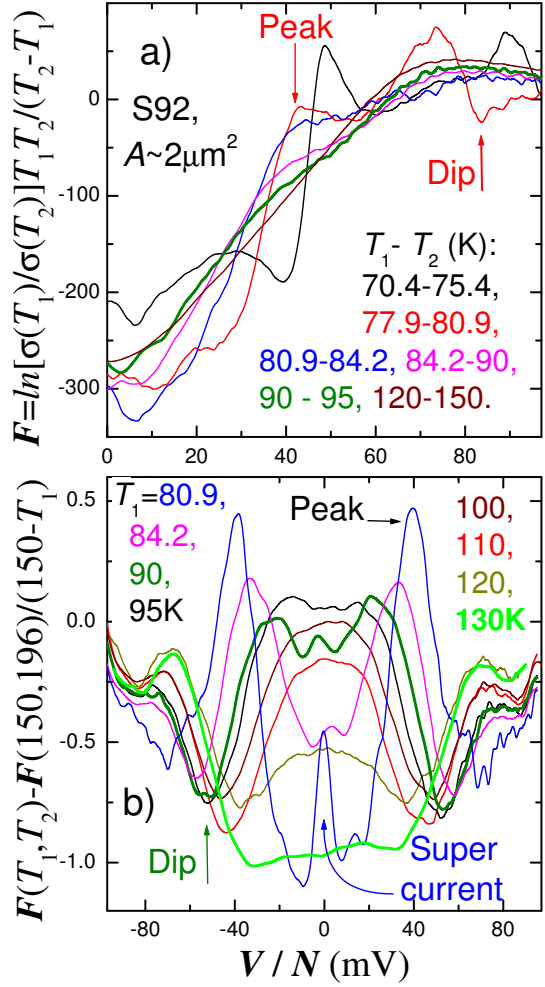}
\caption{\label{Fig7} (Color online). Improvement of spectroscopic resolution at elevated $T$ by means of $T-$differential spectroscopy for the same small mesa on S92 crystal. In a) shown are differences between $ln(\sigma=dI/dV)$ curves from Fig.4 b) at two consecutive $T$, normalized by the temperature difference. Such curves emphasize $T-$dependent spectroscopic features and collapse into the universal curve for the case of TA. In b) the universal TA curve at high $T_1,T_2$ was subtracted to remove completely the TA background. This allows clear observation of evolution of the sum-gap peak and the double-gap dip well above the phase-coherent $T_c^{phase}$.}
\end{figure}

As seen from Figs. 4 and 5 b), the sum-gap peak in $dI/dV$ is rapidly smearing out with approaching $T_c$. Remaining weak spectroscopic features can be traced in a standard way by studying higher derivatives, e.g. $d^2I/dV^2$ \cite{Cascade}. However, they are obscured by the parasitic TA-background. In Ref.\cite{Doping} it was shown that the ITS resolution at $T\lesssim T_c$ can be improved by subtracting the TA curve at $T>T_c$. But since TA is strongly $T-$dependent, such subtraction does not completely remove the changing TA background.

Here I suggest the following optimization for TA background cancelation: First, consider a normalized difference between two characteristics as in Figs. 4,5 b) at nearby temperatures $T_2>T_1$: $F(T_1,T_2)=[ln(dI/dV(T_1))-ln(dI/dV(T_2))] T_1 T_2 /(T_2-T_1)$. According to Eq.(\ref{TeffN}), for pure TA $F(T_1,T_2) \simeq U_{TA} - eV/2$, i.e., is approximately $T-$independent, thus allowing optimal cancelation of the TA background. Another important advantage of this $T-$differential scheme is that it emphasizes any $T-$dependent spectroscopic feature.

In Figs. 7 a) and 8 a) $T-$differential characteristics $F(T_1,T_2)$ are shown for moderately underdoped crystals S92 and S82, respectively. It is seen that substantially above $T_c$ the curves collapse into a single universal curve, as expected for pure TA. At lower $T$, the sum-gap peak and the double-gap dip are clearly resolved up to $T_c$. Interestingly, small deviations from the universal TA curve are seen even above $T_c$ up to $\sim 130K$. To see them more clearly, in Figs. 7 b) and 8 b) the universal TA curves at high $T_{1,2}$ were subtracted from the $T-$differential characteristics. In such the plot the TA background is completely removed and we can very clearly see the remaining spectroscopic features close and even above $T_c$.

\begin{figure}
\includegraphics[width=3.0in]{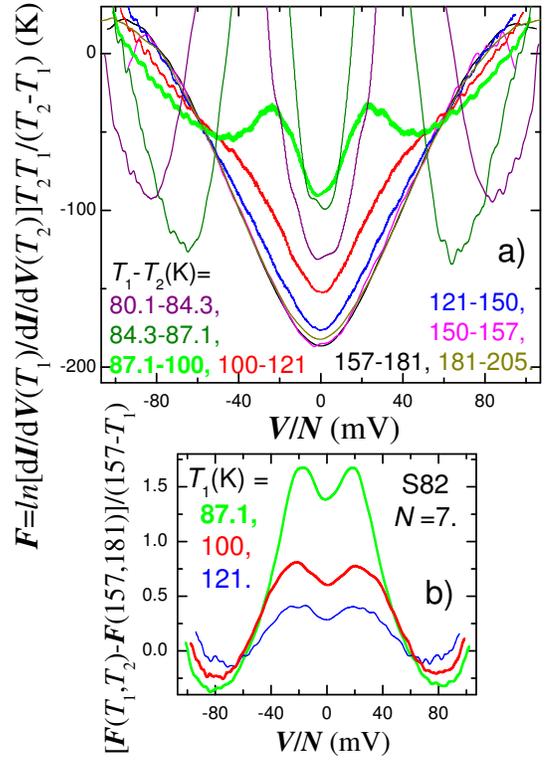}
\caption{\label{Fig10} (Color online). The same as in Fig. 7 for another underdoped S82 crystal. The collapse of curves at high $T$ into the universal TA curve is clearly seen in a). From panel b) it is seen that the normal state pseudogap remains roughly $T-$independent at $T>T_c$.}
\end{figure}

\section{Temperature dependence of the bulk energy gap}

%Thus, self-heating can be effectively obviated by miniaturization of mesas. Moderate self-heating can be easily accounted for and compensated, as will be shown below. Thus unambiguous spectroscopic information about bulk energy gap in HTSC can be recovered.

Figs. 9-12 summarize temperature dependencies of ITS features for different samples. Data for the same mesas are represented by symbols of the same type and color. 

Dashed-dotted lines represent the TA barrier, obtained from the zero-bias resistance using Eq.(\ref{TeffN}) $U_{TA}=k_B T \ln(R_0/T)$. It is seen that $U_{TA}$ is practically constant at $T>T_c$. The sudden fall of $U_{TA}$ marks the superconducting transition.%, which starts at the onset temperature $T_c^{onset}$ and ends at the phase-coherent temperature $T_c^{phase}$, at which the detectable $c-$axis critical current appears, see Table-I. The detailed view of transition for near optimally doped crystals is shown in Fig. 13.

Crosses in Figs. 10-12a) represent crossing points \cite{Sven}, marked in Figs. 2 and 4b). In agreement with Fig. 6 they occur at $eV/N \sim 2U_{TA}$. 

Open symbols in Figs. 9-11 a) show the hump voltage \cite{KrTemp}. It also represents the TA barrier and is roughly $T-$independent at $T>T_c$ \cite{KrTemp,Doping}. The hump appears at slightly higher voltage than $2U_{TA}/e$, also in agreement with Fig. 6. 

From Figs. 10-12 a) it is seen that in most underdoped crystals a slight deviation of $U_{TA}$ downwards, and the crossing point upwards occur below some temperature $T^* \sim 130-150 K$. Dashed and dotted vertical lines mark the $T_c^{phase}$ and $T^*$, respectively.

Solid symbols in Figs. 9-12 a) represent the main experimental result of this work: $T-$ dependencies of sum-gap voltages $eV_{sg}/N=2\Delta$. %, for mesas with consecutively decreasing doping.
Data points were obtained from the peak and half the double-gap dip in $dI/dV$ (larger symbols), or $T-$ differential characteristics (smaller symbols). The latter allows us to trace the gap up to considerably higher temperatures, than before.

In agreement with previous reports \cite{KrTemp,Doping} the bulk gap considerably decreases at $T\rightarrow T_c$ for all doping levels. Simultaneously, we observe that the hump also becomes $T-$dependent at $T<T_c$ \cite{Doping}. However, it moves approximately two times slower than the sum-gap peak and is approximately described by the expression $eV_{hump}(T<T_c) \simeq 2U_{TA}+\Delta(T)$. This indicates that the $c-$axis TA barrier remains intact by the superconducting transition \cite{KrTemp} and continue to hinder the QP transport at $T<T_c$.

\begin{figure}
\includegraphics[width=3.0in]{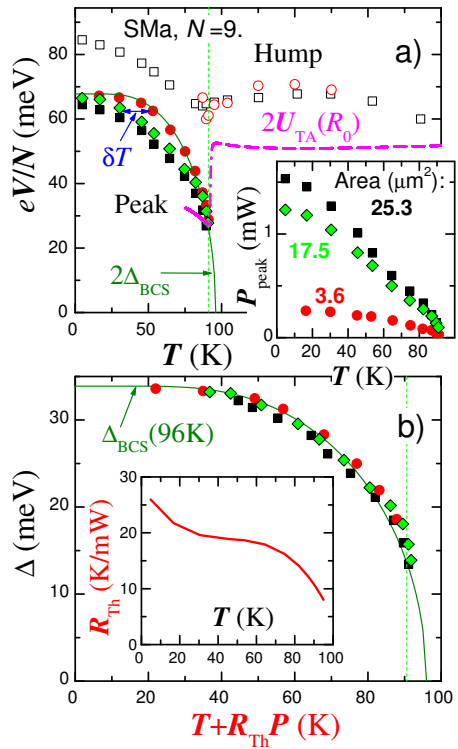}
\caption{\label{Fig9} (Color online).
a) Temperature dependence of characteristic spectroscopic features for three mesas on the near optimally doped SMa crystal. Data for the same mesas are shown by symbols of the same type and color. Open symbols - the TA hump, the dashed-dotted curve - $2U_{TA}$. Downturn of $U_{TA}$ marks the $c-$axis resistive transition, $T_c^{phase}$ at which $c-$axis phase coherence is established is marked by the dashed vertical line. Solid symbols - the sum-gap peak,
it is seen that the measured gap is distorted with increasing mesa area exactly as expected for self-heating. Inset in a) shows power at the peak for the same mesas. b) The heat-compensated, $T-$dependence of the bulk energy gap. Solid line shows that it is very well described by the conventional BCS temperature dependence with the mean-field critical temperature $T_c^{mf}=96K$. No pseudogap is observed above $T_c^{mf}$. Inset in b) shows the effective thermal resistance for the two larger mesas.}
\end{figure}

\subsection{Size dependence}

Size-dependence of ITS unambiguously reveals the extent of self-heat distortion \cite{Heat}. Such data is presented in Figs. 9 and 12.
Dissipation powers at the sum-gap peak $P_{peak}$ for all studied mesas are shown in insets of Figs. 9 a) and 10-12 b). $P_{peak}$ scales with the mesa area.

Solid symbols in Fig. 9 a) show measured $V_{sg}$ for three mesas on the SMa crystal. Size dependence of ITS for this crystal was reported in Ref.\cite{Heat}. It was shown that for mesas with $A < 15 \mu m^2$ the measured gap becomes size-independent. Therefore, data for the smallest mesa with $A\simeq 3.6 \mu m^2$ represents the genuine, undistorted bulk gap $\Delta(T)$, as concluded in sec. III A. The solid line in Fig. 9 a) shows that it is very well described by the conventional mean-field BCS temperature dependence $\Delta_{BCS}(T)$. The same is true for the small mesa on the S92 crystal in Fig. 10a), which was also identified as heating-free in sec. III A.

Let's now consider the two larger mesas in Fig. 9a). It is seen that measured gaps become progressively smaller with increasing $A$ and $P_{peak}$. The observed deviation from the genuine $\Delta(T)$ is perfectly consistent with self-heat distortion, as shown in Fig. 15 d). The temperature rise $\delta T$ is equal to the horizontal shift of the measured gap with respect to the undistorted $\Delta(T)$, as indicated in Fig. 9 a).

Now we can directly calculate the effective thermal resistance of the mesas: %using Eq.(\ref{EqSHeat}): 
$R_{Th}(T)=\delta T/P_{peak}(T)$. The obtained $R_{Th}(T)$ appeared to be approximately the same for both mesas and is shown in the inset of Fig. 9 b). The values of $R_{Th}$ are ranging from $\sim 25 K/mW$ at $4.2K$ to $\sim 10 K/mW$ at $T_c$, consistent with direct in-situ measurements in Ref. \cite{Heat}. Note that %here we measure the actual thermal resistance of the mesas, while
in Ref.\cite{Heat} a separate thermometer was employed for measuring the mesa temperature. Therefore, agreement in obtained $R_{Th}(T)$ in both cases indicates that there is no major thermal gradient along the crystal near the mesa. %, which would otherwise cause a certain lag between the thermometer and the heated mesa, resulting in some underestimation of $R_{Th}$ in the experiment of Ref.\cite{Heat}. The lack of major thermal gradient and the small value of $R_{Th}$ is inconsistent with heat diffusion model (see e.g. Fig.1 in Ref. \cite{HeatJAP}) and supports the 
This is consistent with the conclusion of Ref. \cite{Heat} that heat transport from the mesa is dominated by ballistic flow of non-equilibrium phonons. %, created by spontaneous emission upon relaxation of tunneled QPs \cite{Cascade}. 
In this case the actual heating starts only deep inside the crystal, $\sim$ phononic mean-free path below the mesa, making both heating and in-plane thermal gradient in the mesa small.

\begin{figure}
\includegraphics[width=3.0in]{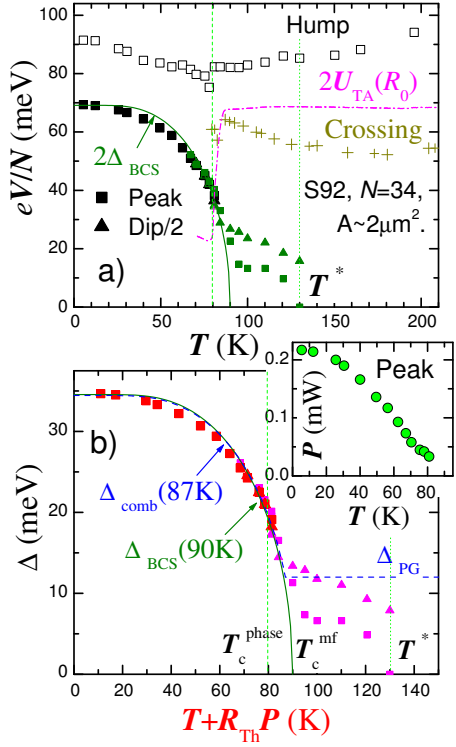}
\caption{\label{Fig10} (Color online). The same as in Fig. 9 for the small mesa on the moderately under S92 crystal. Crosses in a) represent the crossing point of $dI/dV(V)$ curves at two consecutive temperatures, see Fig. 4 b). Filled triangles represent half the double-gap dip. It is seen that signatures of the energy gap (the pseudogap) survive up to $T^*\simeq 130K$ (see Fig.7). Solid and dashed lines in b) show that $\Delta(T)$ is equally well described by the pure BCS-dependence or the combined gap between the BCS gap and $T-$independent pseudogap, shown by the dashed horizontal line. However, in both case the mean-field superconducting temperature is smaller than that for the optimally doped crystal in Fig. 9. }
\end{figure}

\subsection{Self-heat compensation}

The knowledge of the thermal resistance allow us to recover the genuine temperature dependence of the gap even for moderately large mesas. In Fig. 9 b) gap values for all three mesas from SMa are plotted as a function of the actual mesa temperature $T_{corr} = T + P_{peak}(T) R_{Th}(T)$, with $P_{peak}(T)$ and $R_{Th}(T)$ from insets in Figs. 9 a) and b), respectively. It is seen that the genuine $T-$ dependence of the gap is recovered also for larger mesas after such self-heat compensation. The solid line in Fig. 9 b) shows that the genuine $\Delta(T)$ is perfectly described by the BCS dependence with $T_c^{mf}=96K$, which i approximately equal to the optimal $T_c$ of our Bi(Y)-2212 crystals. It is close to $T_c^{onset}\simeq 95K$ and slightly higher than the phase-coherent $T_c \simeq 91.5K$ for this crystal, as shown in Fig. 13.

\begin{figure}
\includegraphics[width=3.0in]{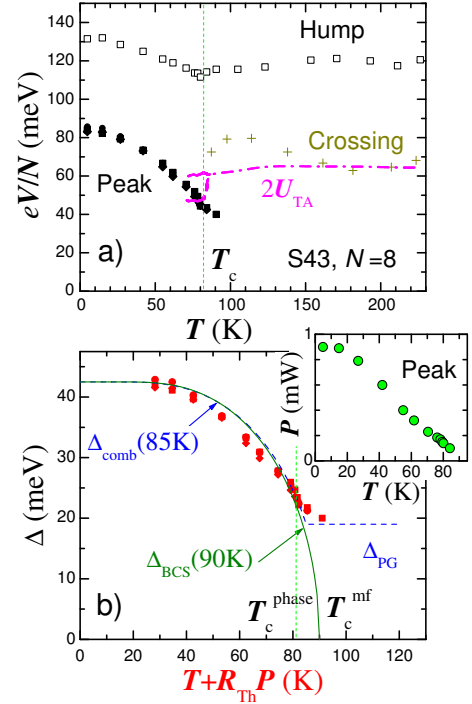}
\caption{\label{Fig11} (Color online). The same as in Fig. 10 for the S43 crystal. It is seen that the superconducting transition at $T_c$ even in moderately underdoped HTSC has a mean-field character and is accompanied by opening of the superconducting gap. }
\end{figure}

We can also check the self-consistency of the conclusion that ITS characteristics of the smallest mesas on SMa and S92 are not distorted by self-heating. As seen from insets in Figs. 9 a) and 10b), the maximum $P_{peak}$ for those mesas is $\sim 0.2 mW$ at the lowest $T$. Therefore, at the lowest $T$ the maximum self-heating $\delta T$ does not exceed few K, which does not affect the measured gap. $P_{peak}$ rapidly decreases with increasing $T$. At $T_c/2$, $P_{peak}$ reduces by half to $\sim 0.1 mW$ and $\delta T \sim 2 K$. Close to $T_c$, self-heating becomes negligible even for moderately large mesas. For the smallest mesas from Figs. 9 and 10, $P_{peak}(T \simeq T_c) \simeq 30 \mu W$ and $\delta T$ is sub-Kelvin. Therefore, the measured gaps for the two smallest mesas are indeed  undistorted by self-heating.

Figs. 10-12 b) represent self-heat compensated $\Delta(T)$ for moderately underdoped mesas. The same $R_{Th}(T)$ %, shown in the inset of Fig. 9, 
was used for self-heat compensation. In all cases, the recovered $\Delta(T)$ can be very well described by the mean-field BCS dependence (solid lines) with $T_c^{mf}$ that is larger than $T_c^{Phase}$ and close to $T_c^{onset}$. Those values are summarized in Table-I.

\begin{figure}
\includegraphics[width=3.0in]{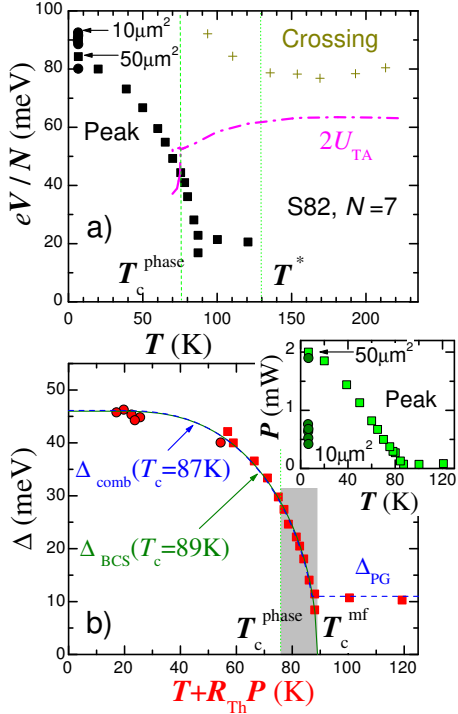}
\caption{\label{Fig12} (Color online). The same as in Fig.10 for the S82 crystal. Data for mesas with different area at $T\simeq 6K$ are shown. Gray area in b) highlights the $T-$region without $c-$axis phase coherence but with persisting gap and in-plane superconductivity. }
\end{figure}

\subsection{Phase-coherence in $c-$axis and $ab-$plane}

In recent years there were many speculations about persistence of phase-incoherent superconductivity up to very high temperatures above $T_c$ in underdoped HTSC. Therefore, it is necessary to clarify the definition of $T_c$ and the difference between $c-$axis and $ab-$plane phase coherence.

The $c-$axis phase coherence is caused by weak Josephson coupling between CuO planes: $E_J=(\hbar /2e)I_c$, where $I_c$ is the total Josephson critical current of the junction. The $I_c$ and the Josephson coupling can be easily suppressed by small magnetic fields, which do not affect superconductivity of planes. Moreover, the Josephson coupling can be simply reduced by reducing the junction area. At $T \gtrsim T_J=E_J/k_B \simeq 23.8 K (I_c/\mu A)$, thermal fluctuations destroy phase coherence and the junction enter in the phase diffusion state with non-zero resistance at zero current \cite{Vion,PhaseDiff}. Suppression of the critical current by thermal fluctuations at $T \ll T_c$ was indeed observed in small mesas \cite{Fluct,PhaseDiff}. Therefore, the $c-$axis phase coherence in small Bi-2212 mesas is not a material property and should not be confused with a much more robust in-plane phase coherence.

\begin{figure}
\includegraphics[width=3.0in]{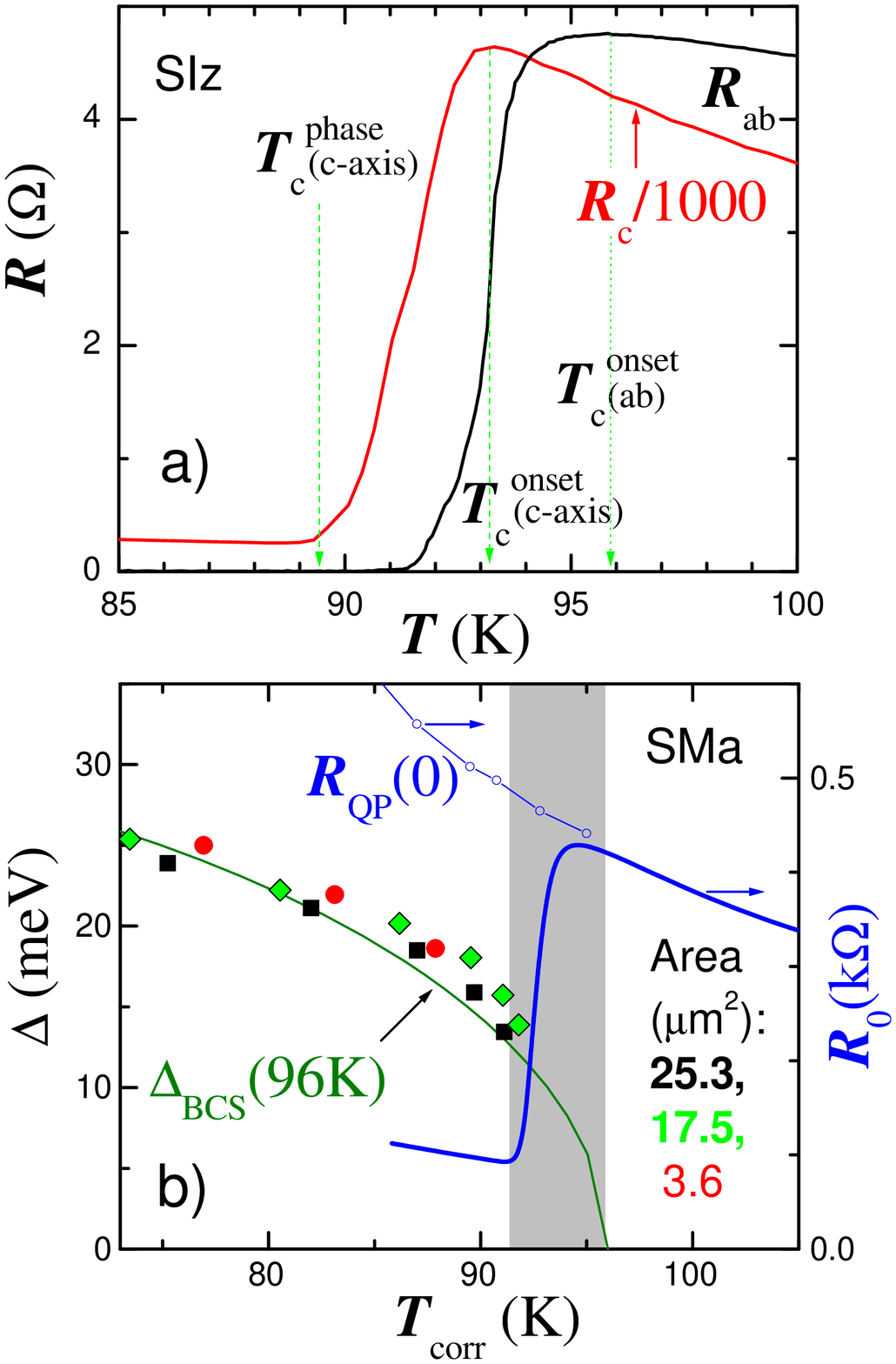}
\caption{\label{Fig13} (Color online).
a) In-plane $R_{ab}$ and $c-$axis $R_c$ resistive transitions in a slightly underdoped crystal, similar to SMa. Panel b) reproduces $\Delta (T)$ from Fig. 9 b) along with the resistive transition for a mesa on SMa. It is seen that the onset of in-plane superconductivity coinsides with the extrapolated mean-field critical temperature, the onset of $c-$axis transition is close to the middle of the in-plane resistive transition, while the temperature at which $c-$axis phase coherence is established is further reduced due to thermal fluctuations. Gray area in b) marks the superconducting region without phase coherence in the $c-$axis direction. }
\end{figure}

To clarify the difference between in-plane and $c-$axis phase coherence, in Fig. 13 a) I show details of in-plane and $c-$axis resistive transitions, measured on a slightly underdoped crystal SIz, similar to SMa. It is seen that the $c-$axis phase coherence is established at $T_c^{phase} \simeq 89.5K$. The onset of $c-$axis transition occurs at $T_c^{onset}(c-axis) \simeq 93K$, which is close to the middle point of the in-plane transition. The onset of $ab-$plane resistive transition occurs at $T_c^{onset}(ab) \simeq 96K$, which coincides with the mean-field critical temperature $T_c^{mf}\simeq 96K$, obtained by extrapolation of $\Delta(T)$ using BCS temperature dependence, as shown in Fig. 13 b)

\subsection{The pseudogap}

The main difference between near optimally doped (Fig. 9) and moderately underdoped (Figs. 10-12) crystals is the persistence of some signature of the residual energy gap (the pseudogap $\Delta_{PG}$) at $T>T_c$ up to $T^* \sim 130-150K$ (see Figs. 7, 8). In the same range $T_c<T<T^*$ we observe an upturn of the crossing voltage and downturn of $U_{TA}$ obtained from zero bias resistance. Thus the pseudogap is seen at all bias levels.

From Figs. 10-12 b) it is seen that the PG is almost $T-$independent near $T_c^{mf}$ in contrast to the strong $\Delta(T)$ dependence on the superconducting side. It is also seen that the PG is merging with the superconducting gap at $T_c$. The overall $T-$dependence of the gap both above and below $T_c$ is well described by the combined gap expression:

\begin{equation}\label{Dcomb}
\Delta_{comb}(T) =\sqrt{\Delta_{BCS}(T)^2+\Delta_{PG}^2},
\end{equation}

with constant $\Delta_{PG}$. Fits of Eq.(\ref{Dcomb}) to experimental $\Delta (T)$ are shown by dashed blue lines in Figs. 10-12 b). Below $T_c$ they are equally good as BCS fits without the PG, but assume lower $T_c^{mf}$, indicated in the figures and Table-I.

Can this pseudogap be related to superconductivity? The answer is rather straightforward. %Whether or not the observed pseudogap represent a true order parameter is discussable. However,
The very fact that it forms the combined gap with the superconducting gap clearly indicates that it represents another order parameter, co-existing and competing with superconductivity.

Our present data, together with earlier observation of coexistence of the superconducting gap and the pseudogap at $T<T_c$ \cite{KrTemp}, are consistent with recent ARPES experiments \cite{Lee2007}, which demonstrated that the superconducting gap along the Fermi arc closes at $T_c$ at all doping levels, while the gap along antinodal directions (the pseudogap) remains relatively $T-$independent at $T_c$.

\section{Discussion and conclusions}

I have shown that technical problems of Intrinsic Tunneling Spectroscopy, such as self-heating (see Appendix-A) and in-plane resistance (see Appendix-B), can be effectively obviated by reducing mesa size. It is instructive to remind that exactly the same technical problems were encountered at the early stage of experimental studies of conventional LTSC tunnel junctions \cite{Giaever4p} and were also solved by junctions miniaturization, which reduces the area-to-perimeter ratio of the junctions. 

I have also shown that self-heating is a trivial phenomenon. %(the simplest problem as far as HTSC spectroscopy is concerned). 
It can be easily accounted for and, in case of moderate heating, compensated, so that the genuine $\Delta(T)$ can be recovered. A rule of thumb for ``moderate" self-heating is that it should not be obvious in the IVCs: the QP branches should reman periodic and there should be no back-bending at $V_{sg}$. All mesas studied here fall in this category.
 
The main result of this work is the uncovered genuine temperature dependence of the bulk energy gap in Bi-2212. For all studied doping levels, $\Delta(T)$ exhibits a strong $T-$dependence, and the superconducting part of it unmistakably tends to vanish in the mean-field BCS manner. For slightly overdoped Bi(Y)-2212 crystals from the same batch this was shown in Ref.\cite{Doping}. Here I have focused on the underdoped side and have shown that the gap vanishes {\it exactly} in the BCS manner also in slightly underdoped (Fig. 9b) and moderately underdoped (Figs. 10-12) mesas.

Those results are %consistent with recent ARPES measurements \cite{Lee2007}, but 
strikingly different from the complete $T-$independent surface gap, reported in STM experiments \cite{Renner}. I want to emphasize that this discrepancy can not be attributed to self-heating and must find another explanation. Indeed, numerical simulations in Fig. 15 clearly show that the trivial self-heating simply can not ``hide" the qualitative $\Delta(T)$ dependence. For example, there is no way in which one can get the vanishing ``measured" (self-heating affected) gap if the true gap is $T-$independent. Furthermore, self-heating becomes insignificant at elevated $T$ because $P_{peak}(T\rightarrow T_c^{mf})\rightarrow 0$. On the other hand, STM characteristics behave similar to the $T-$independent hump feature in ITS, which is the consequence of the $c-$axis thermal activation barrier. As discussed in sec. III-B,  $U_{TA}$ is most likely the property of the blocking BiO plane, which is probed by STM, rather than superconducting CuO planes.  %As shown in sec. II-B, electronic properties of Bi-2212 change dramatically at the scale of just one atomic layer.
The dramatic difference between STM spectra on  BiO and CuO surfaces was indeed reported \cite{Misra}. This highlights the significance of spectroscopic information from bulk CuO planes, obtained here.

Main conclusions, that can be drawn from ITS data are: 

(i) The superconducting transition, even at moderate underdoping, is due to conventional mean-field phase transition rather than destruction of phase coherence without amplitude fluctuations.
This confronts speculations about persistence of the ``precursor" superconducting state in the extended $T-$region above $T_c$.

(ii) The mean-field superconducting critical temperature decreases with underdoping. Thus, HTSC does not become stronger with approaching the undoped antiferromagnetic state.

(iii) The pseudogap co-exists and competes with superconductivity and disappears at optimal doping.

An important consequence of those conclusions is that high temperature superconductivity is strongest at optimal doping and becoming weaker with underdoping. This is consistent with the decrease of the upper critical field \cite{Hc2} with underdoping. Therefore, our observations support the idea that the mechanism of HTSC is intimately connected to a Quantum Critical Point near optimal doping \cite{TallonPhC,QCP}, rather than closeness to the antiferromagnetic state \cite{AF}.

\begin{acknowledgments}
I am grateful to A.Yurgens, for providing
Bi(Y)-2212 crystals, to I.Zogaj, M.Sandberg, and A.Yurgens for assistance at early stage of this work. Financial support from the K.{\&}A. Wallenberg foundation and the Swedish Research Council is gratefully acknowledged.
\end{acknowledgments}

\appendix
\section{Analysis of self-heating in ITS}

Despite relative simplicity of self-heating  phenomenon (it is certainly the most trivial problem in HTSC spectroscopy), discussion of self-heating in ITS has caused a considerable confusion, a large part of which has been caused by a series of publications by V.Zavaritsky \cite{Zavar}, in which he ``explained" the non-linearity of ITS characteristics by assuming that there is no intrinsic Josephson effect. Irrelevance of that model for ITS was discussed in Refs. \cite{HeatCom}.

A certain confusion might be also caused by a large spread in thermal resistances, $R_{Th}$, reported by different groups \cite{Gough,AYreply,Heat,WangH}. For the sake of clarity it should be emphasized that those measurements were made on samples of different geometries. It is clear that $R_{Th}$ depends strongly on the geometry \cite{HeatJAP,Heat}, and is much larger in suspended junctions with poor thermal link to the substrate \cite{WangH} than in the case when both top and bottom surfaces of the junctions are well thermally anchored to the heat bath \cite{Lee}.
For mesa structures similar to those used in this study (a few $\mu m$ in-plane size, containing $N\simeq 10$ IJJ), there is a consensus that $R_{Th}(4.2K) \sim 30-70 K/mW$ (depending on bias) \cite{AYreply,Heat} and $R_{Th}(90K)\sim 5-10 K/mW$ \cite{Heat}. Larger values $R_{Th} > 100K/mW$ claimed by some authors \cite{Gough} are unrealistic for our mesas because they can withstand dissipated powers in excess of $10 mW$ without being melted.

Yet, talking about a typical value of $R_{Th}$ is equally senseless as talking about a typical value of a contact (Maxwell) electrical resistance: both depend on the geometry. Therefore, reduction of mesa sizes provides a simple way for reduction of self-heating \cite{HeatJAP}. Consequently, variation of $dI/dV$ characteristics with the junction size and geometry provides an unambiguous way of discriminating artifacts of self-heating from the spectroscopic features \cite{Heat}.

\subsection{Peak splitting in non-uniform junctions}

%An alternative interpretation of the peak in $dI/dV$ was proposed in Ref.\cite{Zavar}, where it was attributed to acute self-heating of the mesa. The proposed model was inappropriate for IJJs because it neglects the very presence of the intrinsic Josephson effect and was re-footed in Refs.\cite{Heat,HeatCom}. Here I provide an additional simple argument against self-heating origin of the peak in $dI/dV$.
As discussed in Ref.\cite{HeatCom}, atomic separation between IJJs precludes any substantial temperature difference between them. Thus, all junctions in a mesa warm up synchronously and there may be only one collective artefact of heating for all IJJs in the mesa. To the contrary, if the peak is the sum-gap singularity - it is an individual property of each IJJ. If junctions are not perfectly identical, the peak in $dI/dV$ will split in up to $N$ sub-peaks. Indeed, peak splitting is quite often observed in experiment and was already reported in Ref.\cite{KrPhysC} along with supporting numerical simulations.

%\begin{figure}
%\includegraphics[width=2.8in]{F7_NonUniform_dIdV_S512.EPS}
%\caption{\label{Fig1} (Color online).a)dIdV S512 b) numerical simulations. Data from Ref.\cite{KrPhysC}}
%\end{figure}
\begin{figure}
\includegraphics[width=2.8in]{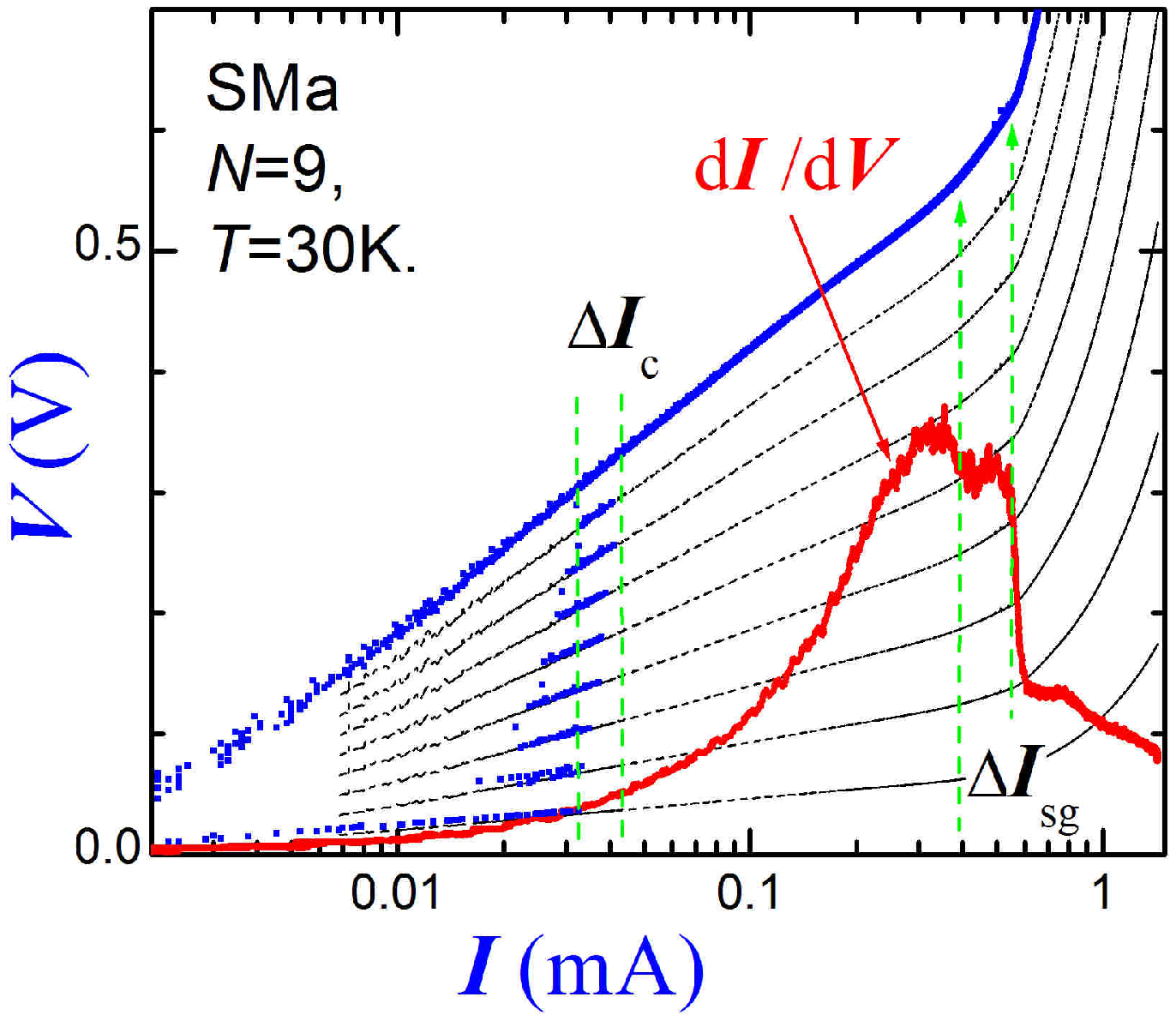}
\caption{\label{Fig14} (Color online). The IVC at $T=30K$ from Fig.2 in a semi-log scale. Thin lines are multiple integers, indicating good periodicity of quasiparticle branches. However, a small nonuniformity of the junctions is seen from the spread of critical currents $\Delta I_c$, which leads to splitting $\Delta I_{sg}$ of the sum-gap peak in $dI/dV (I)$ (red line) because different junctions reach $V_{sg}$ at slightly different currents.}
\end{figure}

In Fig.14 the IVC at $T=30K$ from Fig.2 is re-plotted in a semi-logarithmic scale $V$ vs $\ln I$. As discussed in Ref.\cite{Sven} the IVC becomes nearly linear on such a scale. Thin lines are multiple integers of the last branch divided by $N=9$. Coincidence of those with quasiparticle branches indicates good periodicity of the latter. However, a minor non-uniformity is observed as a gradual increase of $I_c$ with the brunch number. The total spread of the critical current $\Delta I_c$ from the second to the last IJJ is marked in the Figure. The red line shows $dI/dV(I)$ for the same IVC. A small splitting of the peak $\Delta I_{sg}$ is seen. Thus, the peak is not a collective phenomenon of the whole mesa, but is a genuine characteristic of each individual IJJs. From Fig. 14 it is seen that $\Delta I_{sg}$ has approximately the same width in the logarithmic scale as $\Delta I_c$. Therefore the splitting is proportional to the difference in critical currents of IJJs in the mesa and is due to the corresponding spread in currents $I_{sg}$ at which individual junctions reach $V_{sg}$.

Thus, non-uniformity of junctions, although usually unwanted for ITS, helps to understand the origin of the peak in ITS characteristics of small Bi-2212 mesas.

\subsection{How self-heating affects $I-V$ characteristics.}

How self-heating {\it can} distort the IVC's of Josephson junctions is obvious: since self-heating rises the effective $T$ it may affect the IVC only via $T-$dependent parameters. There are three such parameters:

i) the quasiparticle resistance,

ii) the superconducting switching current,

iii) the superconducting gap.

They will affect the IVC of mesas, containing several stacked IJJs, in the following manner:

The consecutive increase of $T$ upon sequential switching of IJJs from the superconducting to the resistive state will distort the periodicity of quasiparticle branches. Each consecutive QP branch will have a smaller QP resistance (smaller $V$ at given $I$) and smaller switching current. This type of distortion becomes clearly visible (at base $T=4.2K$) when $\Delta T \gtrsim 20K$ \cite{HeatJAP,HeatCom}.

For better understanding of the influence of self-heating on IVCs of Josephson junctions, in Fig. 15 I reproduce the results of numerical simulation of such the distortion, made specifically for the case of Bi-2212 mesa with the corresponding $T-$dependent parameters (see Ref.\cite{KrPhysC} for details). Fig. 15 a) shows a set of undistorted IVC's at different $T$ for coherent, directional, $d-$wave tunneling with some trial $\Delta(T)$, shown by the solid line in Fig. 15 d). Panels b) and c) show the distorted IVCs and the actual junction temperature, respectively. It is seen that combination of self-heating and $T-$dependence of $\Delta$ may lead to appearance of back-bending of the IVC at the sum-gap knee. The dashed line in panel d) represents the ``measured" gap obtained from the peak in distorted $dI/dV$ characteristics. Remarkably, the deviation from the true $\Delta(T)$ is marginal, despite large self-heating, $\Delta T \simeq T_c/2$ at $4.2 K$! Numerical simulations has shown that even self-heating up to $T_c$ at the sum-gap knee does not cause principle changes in the behavior of the ``measured" gap. The robustness of the measured gap with respect to self-heating is due to the flat $T-$dependence of the superconducting gap at $T<T_c/2$ and to simultaneous vanishing of dissipation power at $V_{sg}$ together with $\Delta(T)$ at $T\rightarrow T_c$, as shown in insets of Figs. 9-12.

\begin{figure}
\includegraphics[width=3.5in]{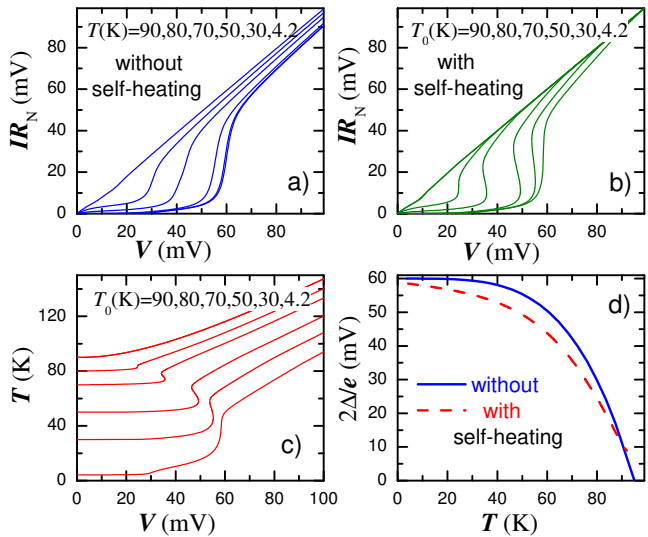}
\caption{\label{Fig15} (Color online). Simulated distortion of SIS tunneling characteristics by strong self-heating.
a) Undistorted IVC's at different $T$. Simulations were made for typical parameters of our mesas, $T-$dependent thermal conductivity of Bi-2212, and for coherent, directional, $d-$wave tunneling. b) distorted IVC's at the same base $T$; c) The mesa temperature as a function of bias. d) $T-$ dependence of the genuine superconducting gap (solid line) and the ``measured" gap obtained from distorted IVC's (dashed curve). Note that even strong self-heating ($T$ reaches $T_c/2$ at $V_{sg}$ at 4.2K) does not cause considerable distortion of the measured gap. Data from Ref.{\cite{KrPhysC}}}
\end{figure}

\subsection{Heating or non-equilibrium phenomena?}

Finally, it is important to emphasize that the concept of heat diffusion is inapplicable for small Bi-2212 mesas containing only few atomic layers. The phonon transport in this case is ballistic \cite{Heat,Uher} and the energy flow from the mesa is determined not by collisions between the tunneled non-equilibrium QP with thermal phonons, but by spontaneous emission of a phonon upon relaxation of the non-equilibrium QP \cite{Cascade}. This process is not hindered at $T=0$. Therefore, the effective $R_{Th}$ (and self-heating) can be much smaller because it is not limited by poor thermal conductivity at $T=0$, but is determined by the fast, almost $T-$independent, non-equilibrium QP relaxation time. The concept of self-heating becomes adequate only in the bulk of the Bi-2212 crystal, where the dissipation power density and the temperature rise are much smaller due to the much larger area of the crystal. For more details see the discussion in Ref. \cite{Heat}. The non-equilibrium energy transfer channel is specific for atomic scale intrinsic Josephson junctions made of perfect single crystals. It can explain a remarkably low self-heating at very high bias \cite{Cascade}.

\section{Artifacts of in-plane resistance}

In experiments on large mesas \cite{Gough} or suspended structures \cite{WangH} no clear Ohmic tunneling resistance could be observed. Instead a negative differential resistance (acute back-bending) progresses at high bias. Similar behavior is also observed on moderately large mesas, when measurements are made in the four-probe configuration \cite{YouPRB}. This is in stark contrast to three-probe measurements on small mesas, reported here, see Fig. 2 and Refs. \cite{KrTemp,KrMag,Doping,Heat}.

The continuous negative resistance is not described by self-heating in tunnel junctions because those should always reach the $T-$independent positive tunnel resistance at high bias. Thus self-heating is not the primary cause of the acute back-bending. Rather, the negative differential resistance, observed in large structures is caused by the loss of equipotentiality of CuO planes upon which the measurement of IVCs in mesa structures is relying. The latter can be triggered by development of a hot spot \cite{HotSpot}, but can even occur without self-heating due to in-plane resistive transition of CuO planes when the applied current exceeds the critical current of the CuO plane \cite{YouPRB}.

\subsection{Acute back-bending without heating and the difference between 3-and 4-probe measurements}

Here I demonstrate how acute, and not recovering back-bending develops in the IVC of the mesa as a result of the finite in-plane resistance and without any self-heating. I also explain the difference between four-probe measurements, that exhibits acute back-bending \cite{YouPRB}, and three-probe measurements - that don't.

\begin{figure}
\includegraphics[width=2.8in]{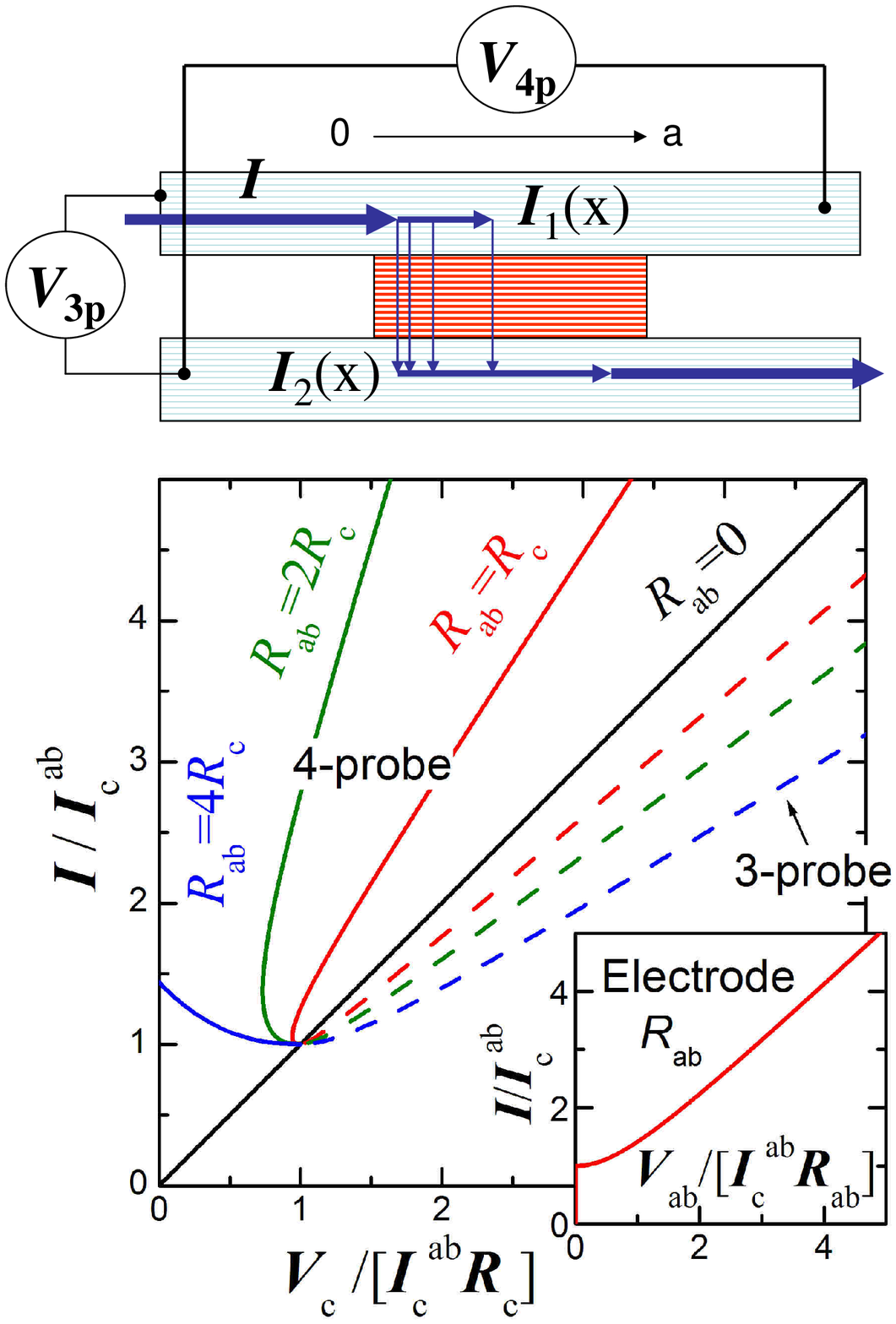}
\caption{\label{Fig16} (Color online). Top panel shows a sketch of current distribution in case of resistive electrodes and contact configuration for three- and four-probe measurements. The main panel shows simulated $I-V$ curves in the three-probe (dashed lines) and four-probe (solid lines) configurations for different in-plane electrode resistances. It is seen that in-plane resistance distorts the measurement of junction characteristics nd that acute back-bending can develope in the four-probe case for high in-plane resistance. Inset show the in-plane IVC.}
\end{figure}

The top panel in Fig.16 show the sketch of the mesa structure. The current $I$ is biased through the top electrode and returned through the crystal, which serves as the bottom electrode. The voltage can be measured either in four-probe $V_{4p}$ or thee-probe $V_{3p}$ configuration, as shown in the Figure.

If electrodes (both top and bottom) are resistive, then there is a voltage gradient along the electrodes and the bias current is distributed non-uniformly within the mesa, as shown by thin vertical arrows. Let $R_{1,2}$ be the in-plane resistance of top and bottom electrodes within the mesa area $0>x>a$, and $R_c$ the $c-$axis resistance of the mesa itself. Following Ref.\cite{Giaever4p}, the current and voltage distribution in this case can be described by the system of equations:

\begin{eqnarray*}
  \frac{dV_1}{dx} = -\frac{R_1}{a}I_1, \\
  \frac{dV_2}{dx} = -\frac{R_2}{a}I_2, \\
  I_1(x)+I_2(x) = I, \\
  V_1(x)-V_2(x) = -R_c a \frac{dI_1}{dx},
\end{eqnarray*}

with the boundary conditions: $I_1(0)=I$ and $I_1(a)=0$.

In the four-probe configuration, the measured voltage is $V_{4p}=V_1(a)-V_2(0)$, which yields:

\begin{equation}\label{EqA4p}
V_{4p}=\frac{I R_1 R_2}{(R_1+R_2)\alpha} \left[ \frac{\left(\frac{R_1}{R_2} +\frac{R_2}{R_1}+2\cosh\alpha\right)}{\sinh\alpha}-\alpha \right],
\end{equation}

where $\alpha=\sqrt{\frac{R_1+R_2}{R_c}}$. In case $R_1=R_2$ it coincides with the result of Ref.\cite{Giaever4p}.

In the three-probe configuration the measured voltage is $V_{3p}=V_1(0)-V_2(0)$, which yields:
\begin{equation}\label{EqA3p}
V_{3p}=\frac{I}{\alpha} \left[ \frac{R_2+R_1\cosh\alpha}{\sinh\alpha}\right],
\end{equation}

The main panel in Fig. 16 shows calculated four-probe (solid lines) and three-probe (dashed lines) IVCs for the case of identical superconducting electrodes $R_1=R_2=R_{ab}(I)$ with the in-plane critical current $I_c^{ab}$ and the IVC as shown in the inset. For $R_{ab}=0$ the measured IVC coincides with real $c-$axis IVC of the mesa. But substantial deviations occur when the bias current exceeds $I_c^{ab}$ and electrodes become resistive.

\subsection{Limitations on the mesa size}

I want to emphasize that large mesas are not suitable for ITS even in the absence of self-heating. To probe the gap, one should be able to reach the sum-gap voltage, $V_{sg}\sim 60 meV$ per IJJ, without loosing the equipotentiality of the bottom CuO plane, which is used as the return current lead and the second voltage electrode. 

Let's estimate the maximum mesa size, suitable for ITS, in the absence of self-heating. Consider a square mesa with the in-plane size $a$. The bias current, required for reaching the sum-gap voltage is $I_{sg}\simeq V_{sg}/(\rho_c s/a^2)$, where $\rho_c \simeq 30 \Omega cm$ is the $c-$axis tunnel (large bias) resistivity, and $s\simeq 1.5 nm$ is the interlayer spacing. This current is flowing through the perimeter of the last IJJ into the bottom CuO layer and should not exceed the in-plane critical current of that layer. Provided the in-plane critical current density is $J_c^{ab}\sim 10^7 A/cm^2$ \cite{YouPRB}, the in plane critical current of the bottom CuO plane through the perimeter of the mesa is $\sim 4J_c^{ab} s a$. Therefore, the mesa size should be smaller than $a_{max} \simeq 4J_c^{ab}s^2\rho_c/V_{sg} \simeq 4.5 \mu m$. Thus, miniaturization is essential for ITS. Otherwise the return current and voltage contacts are no longer equipotential, leading to the negative measured differential resistance.

To cause a substantial distortion, the total in-plane resistance should be larger than the mesa resistance. Let's see if this is the case for Bi-2212. The in-plane resistance of the bottom CuO plane in the square $a\times a$ is:

$R_{ab}(\Box)=\rho_{ab}a/sa \sim 10^{-4}(\Omega cm)/1.5 (nm) =667 (\Omega)$.

The $c-$axis mesa resistance is $R_{c}=\rho_{c}Ns/a^2$ $\sim 30 (\Omega cm) 1.5(nm) N/a^2 = $ $450 N/[a(\mu m)]^2 (\Omega)$.

For a mesa $5\times 5 \mu m^2$ with $N=10$ IJJs, $R_c = 180 \Omega$ is about four times smaller than $R_{ab}(\Box)$. Thus we see that distortion by in-plane resistance can indeed be significant for larger mesas with a small number of junctions.

The proposed model explains why four- and three-probe measurements of Bi-2212 mesas may be very different. From Fig.16 it is seen that four- and three-probe measurements react differently on in-plane resistivity. For the three-probe case, it just leads to appearance of an additional series resistance. But for the four-probe configuration it may lead to development of the acute back-bending, precluding any spectroscopic analysis.

The model can also explain a strange behavior of mesas with very small amount of junctions \cite{YouAPL}. From the estimations above it follows that for a mesa $5\times 5 \mu m^2$ with only one IJJ the in-plane resistance is about 40 times larger than the mesa resistance, which make such mesas extremely prone to distortion by in-plane resistance and, probably, not suitable for ITS.

%Finally I note that the proposed model is conceptually similar to that in Ref.\cite{You??}. However, in contrast to the statement of Ref.\cite{You??}, more detailed treatment of the in-plane resistance shows that acute back-bending does not require self-heating.

%\end{multicols}
\end{document}